\documentclass[12pt,aps,amssymb,amsmath,showkeys]{revtex4}
\usepackage{graphicx}
\usepackage{epstopdf}
\usepackage{multirow}
\usepackage{subfloat}
\usepackage{float}
\usepackage{adjustbox}
\usepackage{caption}
\usepackage{placeins}
\usepackage{comment}
\usepackage{mathtools}
\usepackage[title]{appendix}
\usepackage{amsmath,amssymb}
\newcommand{\beq}{\begin{equation}}
\newcommand{\eeq}{\end{equation}}

\begin{document}
\title{First-passage properties of mortal random walks: ballistic behavior, effective reduction of dimensionality, and scaling functions for hierarchical graphs}
\author{V. Balakrishnan}
\email{vbalki@physics.iitm.ac.in}
\affiliation{Department of Physics, Indian Institute of Technology Madras \\
 Chennai  600 036, India}
\author{E. Abad}
\email{eabad@unex.es }
\affiliation{Departamento de F\'isica Aplicada and Instituto de Computaci\'on Cient\'ifica Avanzada (ICCAEx) \\
	Centro Universitario de M\'erida, Universidad de Extremadura, E-06800 M\'erida, Spain}
\author{Tim Abil}
\email{tim.abilzade@gmail.com}
\affiliation{College of Computing and Digital Media \\ DePaul University, 243 South Wabash, Chicago, Illinois 60604-2301, USA}
\author{John J. Kozak}
\email{kozak@depaul.edu}
\affiliation{Department of Chemistry \\
	DePaul University,
	Chicago, Illinois 6604-6116, USA}
\date{\today}
\begin{abstract}
We consider a mortal random walker on a family of hierarchical graphs in the presence of
some trap sites. The configuration comprising the graph,  the starting point of the walk, and
the locations of the trap sites is taken to be exactly self-similar as one goes from one generation of the
family to the next. Under these circumstances, the total probability that the walker
hits a trap is determined exactly as a function of the single-step survival probability
$q$ of the mortal walker. On the $n^{\rm th}$ generation graph of the family, this
probability is shown to be given by the $n^{\rm th}$ iterate of a certain scaling  function or map $q \rightarrow f(q)$.  The properties of the map then determine, in each case,
 the behavior of the trapping probability,
the mean time to trapping, the temporal scaling factor governing the random walk dimension
on the graph, and other related properties. The formalism is illustrated for the cases of a linear hierarchical lattice and the Sierpinski graphs in $2$ and $3$ Euclidean dimensions.
We find an effective reduction of the random walk dimensionality due to the ballistic behavior of the surviving particles induced by the mortality constraint.
The relevance of this finding for experiments involving travel times of particles in diffusion-decay systems is discussed.

\end{abstract}

\maketitle
\section{Introduction}
\label{intro}

Theories of diffusion-reaction processes have, almost universally, been based on the assumption that the medium on or within which transformation(s) take place is homogeneous, free of imperfections. This is the case for continuum diffusion theories based on Fick's laws, where the theory of (linear or non-linear) partial differential equations can be mobilized \cite{NP1977}. It is also the case for lattice-based theories in which translational invariance is assumed; here, for instance,  the  generating
functions technique can be used to obtain analytic or asymptotically-exact results \cite{W94, H95, KMS73, HK82, AK15, AASK18}. If spatial imperfections are present in the system, or if there are (uniformly or randomly dispersed) competing reaction centers, analytical results using the foregoing  continuum or lattice approaches are more difficult to obtain. However, considerable  progress has been made  in recent years in characterizing space exploration and first-passage properties of random walkers subject to such constraints.  This applies both to systems of standard random walkers and to systems of mortal walkers, i.e., walkers that may die as they walk. Analytic results for both types of walkers have been obtained via different techniques \cite{KS60, K00, WK81, KB02IJBC, PK83, PK85, SMK85, KB02PRE, WK82, BK13PRE, KNN00, NKN01, W94, H95, KMS73, HK82, AK15, AASK18, W64, W69, YAL13, AYL13, YAL14, BR14, CAMBL15, MR15, M15, RB15, OMR16, GR17, BRB17, B17, CBM17, ER17, MW65, MS57, A05, CBM07, BV14, EN85}, e.g., the theory of Markov processes \cite{KS60, K00, WK81, KB02IJBC, PK83, PK85, SMK85, KB02PRE, WK82, BK13PRE, KNN00, NKN01} and generating function approaches \cite{W94, H95, KMS73, HK82, AK15, AASK18, W64, W69, A05, YAL13, AYL13, YAL14, BR14}. With the development of the theory of random walks on fractals \cite{BH00, ZBK85, RT82, RT83, AY98, AY00, A08, MABV12, LWZ10,BTK10}, results in the exact analysis of dynamical processes taking place in inhomogeneous media have emerged. The presence of lacunary regions allowed an exploration of the consequences of breaking the translational symmetry of the host medium on the reaction efficiency. For example, previous works coauthored by two of us took advantage of fractal self-similarity to obtain an exact analytic expression for the mean walk length (or the mean number of time steps to absorption) of a random walker on the Sierpinski gasket \cite{KB02PRE} and on the Sierpinski tower embedded in an arbitrary number $d$ of Euclidean dimensions \cite{KB02IJBC}. These results were obtained for the case where the diffusing species undertook only nearest-neighbor (NN) displacements. Later, it was shown that analytic results could be obtained for Sierpinski graphs embedded in $d \geq 2$ Euclidean dimensions when both nearest-neighbor and next-nearest-neighbor (NNN) jumps are considered \cite{BK13PRE}. Also of importance for the understanding of the reaction dynamics on Sierpinski graphs is the fact that analytic expressions can be obtained for recurrence relations among the eigenvalues of the operator involved in the underlying master equation \cite{BTK10}. In this study, we show that scaling relations can be obtained for the reaction efficiency (as gauged by the mean time to absorption) for the Sierpinski gasket and tower (thus considering explicitly lacunary regions) for the case of mortal walkers, i.e., the case where the diffusing reactant can be deactivated with a certain probability before it encounters a reaction site (represented by a deep trap).  This premature deactivation can be thought of, for instance,  as arising from the action of additional reaction centers competing with the reaction site. Among the examples of processes in which understanding this competition is of importance is light-energy conversion in the photosynthetic antenna system, first studied analytically by Montroll \cite{W69}.
As will be brought out in Section VI, a description in terms of mortal walkers may also be relevant for another system where light-energy conversion to chemical energy takes place, namely crystalline, luminescent nanofibers of poly(di-n-hexylfluorene) (PDHF) in which exciton diffusion  is observed.  A similar description may also be applicable to the study of valley diffusion currents in TDMC quantum heterostructures.  For both of these systems, our results suggest that a device that is able to detect travel times of the generated excitations would measure values that are significantly smaller than those expected from normal diffusion. This is due to the fact that the excitation decay penalizes long trajectories and therefore long travel times.  As a result of this, the effective random walk dimension is strongly reduced.  One of our main goals will be to quantify this reduction in terms of suitable scaling functions.

The plan of the rest of this paper is as follows.  In Sec. II, we give a general formulation of the mortal random walk problem. We start with the case of diffusion on a line, and show that mortality merely introduces an attenuating factor in the form of a decaying exponential. We then consider the backward Kolmogorov equation (BKE) on a graph in the presence of deep traps,
 and show how the trapping probabilities of standard and mortal walkers are related to each other.   Finally, we illustrate the formalism by an application to a linear lattice with a trap at one end, by deriving an explicit solution for  the conditional mean first passage time (MFPT) to the trap (also termed `conditional walk length' in what follows).  In Sec. III, we extend
the formalism to hierarchical lattices, and explicitly deal with the cases of a hierarchical linear lattice, the Sierpinski gasket, and the Sierpinski tower.  In Sec. IV, we extend the theory to the computation of the unconditional walk length, i.e.,  the length of a mortal walk which is terminated either by absorption at a deep trap or by a premature death of the walker before this happens.  The specific case we consider here is that of a Sierpinski gasket (SG) with a deep trap at a corner site. In Sec. V, comparative results between conditional and unconditional walk lengths are given for more complex situations, involving the Sierpinski gasket with two deep traps and the Sierpinski tower with three deep traps.  Finally, in Sec. VI, we discuss the main implications of our findings for experimental systems.

\section{General formulation}
\label{genlformulation}
A `mortal' random walker has a probability $q$ (where  $0 < q < 1$)  of surviving at each time step,
and therefore a non-zero probability $s=1-q$ of demise at each time step.
Standard random walks  correspond to the limiting case $q = 1$.
We consider a mortal  walker executing a discrete-time
Markovian random walk  via equal-probability jumps to nearest-neighbor (NN)
sites  on a finite connected graph, on which  a specified set of sites comprise so-called deep traps.
Once the walker hits any member of the latter  set  for the first time, the walk is over. The
crucial difference between this
situation and the standard random walk is that the total probability
$P_{j}(q)$ of the trapping of a walker starting from an arbitrary non-trap site $j$ is now less than
unity, in general.  Moreover,  $P_{j}(q)$ serves as a basic quantity that carries essential
information regarding the random walk.
 As we shall see, interesting features ensue from this  fact.

\subsection{The continuum limit}
\label{continuumlimit}

Before turning to the main theme of this work,  we may quickly dispose of the
continuum or diffusion limit of a mortal random walk, if only to point out that
this limiting case  runs along lines that may be expected intuitively.
The exercise does, however, yield some pointers to some features of a
general nature.
For definiteness,
consider a mortal random walker on an infinite linear lattice with its sites labeled by the
integers. Let $a$ and $\tau$ denote, respectively, the lattice constant and time step.
Then, if $p_{j}(n)$ is the probability that the walker is at the site $j$ at time $n\tau$,
we have
\beq
p_{j}(n+1) = \tfrac{1}{2} q\,[p_{j-1}(n) + p_{j+1}(n)].
\label{1drw}
\eeq
The only consistent continuum limit of this difference equation is
obtained by  letting $a\rightarrow 0$ and  $\tau \rightarrow 0$, as usual,
as well as   $q \rightarrow 1$, such that
\beq
\lim \, q a^{2}/(2\tau) = D,\;\;\lim\, (1-q)/\tau = \sigma,
\label{difflimit}
\eeq
where $D$ and $\sigma$ are finite, non-zero constants. Replacing
$ja$ by $x$
and $n\tau$ by $t$ in the customary manner, and retaining the symbol
$p(x,t)$ for the positional probability density,   we obtain from Eq. \eqref{1drw}
a diffusion equation with a linear death term:
\beq
\frac{\partial p}{\partial t} = D\,\frac{\partial^{2}p}{\partial x^{2}} - \sigma \,p.
\label{diffusioneqn}
\eeq
The fundamental solution of Eq. \eqref{diffusioneqn} satisfying the initial condition
$p(x,0) = \delta (x-x_{0})$
is given by
\beq
p(x,t) = e^{-\sigma t} (4\pi D t)^{-1/2}\,e^{-(x-x_{0})^{2}/(4Dt)}.
\label{diffeqnfundsoln}
\eeq
This is just the standard Gaussian solution with the extra
exponentially decaying   factor
$e^{-\sigma t}$.
It may be noted that the pre-factor in the solution
is now determined  from the initial condition, i.e.,
the requirement that $p(x,t) \rightarrow \delta (x-x_{0})$ as $t\rightarrow 0$, rather than
the normalization of $p(x,t)$ to unity---which is no longer the case, because
$\int_{-\infty}^{\infty}\!dx\,p(x,t) = e^{-\sigma t}$. \\

As we shall be interested in trapping probabilities and mean first-passage times, we
consider a first passage from the starting point $x_{0}$ to an arbitrary point $x$
($> x_{0}$,  for definiteness). Let $q(t,x\vert x_{0})$
be the corresponding first-passage-time
density.  This quantity is conveniently determined
from the Markovian renewal equation
\beq
p(x_{1}, t \,\vert\,   x_{0}) = \int_{0}^{t}\!dt^{\prime}\,q(t^{\prime}, x \,\vert \,x_{0})\,
p(x_{1}, t-t^{\prime}\,\vert\, x),
\label{renewaleqn}
\eeq
where $x_{0} < x < x_{1}$. With the help of Laplace transforms, we find
\beq
q(t,x\,\vert \,x_{0}) =
\frac{e^{-\sigma t}}{(4\pi D t^{3})^{1/2}}\,(x-x_{0})\,e^{-(x-x_{0})^{2}/(4Dt)}.
\label{continuumfptd}
\eeq
Once again, this is just the standard expression
(a stable or L\'evy distribution in $t$
with exponent $\frac{1}{2}$),
multiplied by the attenuation factor
$e^{-\sigma t}$.  Owing to this factor, the first moment of
$q(t,x\,\vert \,x_{0})$ is finite, rather than divergent.
The mortality of the walker makes  the random walk  non-recurrent, even in one dimension.
The total probability that a first passage from $x_{0}$ to $x$ occurs at all is given by
\beq
\int_{0}^{\infty}\!dt\,q(t,x\,\vert \,x_{0}) =
 \exp\,[- (\sigma/D)^{1/2}\,(x-x_{0})],
\label{continuumtrapprob}
\eeq
which decreases exponentially as the distance $(x-x_{0})$ increases.  Although the
probability of a first passage to $x$ is less than unity, a mean first-passage time
(MFPT) may still be defined. We find
\beq
T(x\,|\,x_{0}) =
\frac{\int_{0}^{\infty}\!dt\,t\,q(t,x\,\vert \,x_{0})}
{\int_{0}^{\infty}\!dt\,q(t,x\,\vert \,x_{0})}
= \frac{(x-x_{0})}{(4 D \sigma)^{1/2}}\,.
\label{continuummfpt}
\eeq
The mean time is therefore directly proportional to the distance to be covered, as in ballistic motion,
with an effective speed $(4D\sigma)^{1/2}$.
It must be remembered, however, that the MFPT is  an average over only
that  fraction of the realizations of the diffusion process in which a first passage
from $x_{0}$ to $x$ occurs at all, and that this fraction decreases exponentially
as the distance $(x-x_{0})$ increases.

\subsection{Random walk on a graph with traps}
\label{standardcase}

Turning now to Markovian random walks on graphs in discrete time in the presence of
trap sites,
it is helpful to begin with  the standard  case ($q= 1$), in order to
bring out more clearly the
differences that arise when $q < 1$.
Let $\phi_{j}(t)$ be the probability that a walker starting from
any non-trap site  $j$ hits any of the traps for the
first time at discrete time $t$, i.e., the first-passage time distribution for the site $j$.
 Then the backward Kolmogorov equation  (BKE) for $\phi_{j}(t)$ is
\beq
\phi_{j}(t+1) = (1/\nu_j) \sum_{k} \delta_{\langle jk \rangle}\,\phi_{k}(t),
\label{bke1}
\eeq
where  $\nu_{j}$ is the number of nearest neighbors of the  site $j$. The symbol
$\delta_{\langle jk \rangle} $ is equal to
$1$ if $j$ and $k$ are NN sites, and is equal to $0$ otherwise.
By definition,  $\phi_{j}(t) = \delta_{t, 0}$ for each trap site $j$
(the  `boundary conditions'). Likewise,
when $j$ is a non-trap site, $\phi_{j}(0) = 0$ (the initial conditions).
Let $\Phi_{j} = \sum_{t=0}^{\infty} \phi_{j}(t)$ be the total probability
that a walker starting at site $j$ ever reaches a trap. Summing over $t$,
Eq. \eqref{bke1} gives
\beq
\Phi_{j} = (1/\nu_j) \sum_{k} \delta_{\langle jk \rangle}\,\Phi_{k}.
\label{suretrapping1}
\eeq
But if $k$ is a {\em trap} site, then
$\Phi_{k} =
\sum_{t=0}^{\infty} \delta_{t,0} =1$. Because of this fact,
\eqref{suretrapping1} becomes an {\em inhomogeneous}
 equation whenever $j$ has a nearest-neighbor trap site.
Equation \eqref{suretrapping1}, written down for every value of $j$,
yields  a
set of linear simultaneous equations with a non-vanishing
discriminant.  Hence there is a unique solution set, deduced by
inspection to be simply  $\Phi_{j}= 1$ for every $j$.
In other words, trapping is a {\it sure}  event  for random walks on all the finite connected
structures in which we are interested, i.e., the first-passage time distribution
$\phi_{j}(t)$ is properly normalized:
\beq
\Phi_{j} = \sum_{t=0}^{\infty}\phi_{j}(t) = 1.
\label{suretrapping2}
\eeq
 (In the case of fractal graphs,
this remains true  in the infinite generation limit, since we only consider
cases in which the
the spectral dimension  $< 2$.)

The `local mean value'  nature of $\phi_{j}(t)$ is evident in Eq. \eqref{bke1}.
This  may be  made more manifest by re-writing it in the form
\beq
\phi_{j}(t+1) - \phi_{j}(t) =  \sum_{k}\Delta_{jk}\,\phi_{k}(t),
 \label{bke2}
 \eeq
where
\beq
\Delta_{jk} = \nu_{j}^{-1} \delta_{\langle jk \rangle} - \delta_{jk}
\label{discretelap}
\eeq
is
(a component of) the discrete Laplacian. The MFPT, or the mean time to trapping
for walks originating from any given site $i$, is the first moment of
$\phi_{j}(t)$, and is defined as
\beq
T_{j} = \sum_{t=0}^{\infty} t\,\phi_{j}(t){\Big/}
\sum_{t=0}^{\infty} \phi_{j}(t)
= \sum_{t=0}^{\infty} t\,\phi_{j}(t),
\label{usualmfptdefn}
\eeq
in view of the normalization in Eq. \eqref{suretrapping2}.
Multiplying both sides of Eq. \eqref{bke1} by $(t+1)$ and summing over $t$
leads to the set of linear simultaneous equations for $\{T_{j}\}$
given by
\beq
\sum_{k}\Delta_{jk}\,T_{k} = -1.
\label{usualmfptbke}
\eeq
It is important to bear in mind (for what follows) that
such a linear relationship among the MFPTs from
different sites is only possible because the distribution $\phi_{j}(t)$ is normalized to unity
for every $j$. Depending on the structure of the hierarchical and/or fractal  graph
concerned, and its symmetries
in the presence of the deep traps, various scaling relations for
partial sums of the MFPTs arise from the appropriate application of Eq. \eqref{usualmfptbke}.
These are relatively simple additive and multiplicative relations, again because of the
linearity of Eq. \eqref{usualmfptbke}.  In particular, they help answer a basic question
related to random walks on hierarchical graphs: if the spatial scaling factor of the graph
that takes us from one generation to the next is $\lambda$,
say, what is the corresponding temporal scaling factor $\mu$?  The ratio
\beq
d_{w} = (\ln\,\mu)/(\ln\,\lambda)
\label{walkdimension}
\eeq
is then the random walk dimension of the hierarchical graph or fractal.
Thus, the fact that, on the average, ``it takes four times as long to go twice as far''
in  conventional diffusion implies that $d_{w} = 2$ in this case.
On the Sierpinski gasket, in contrast,  it is well known that
``it takes (on the average) five times as long to go twice as far'', implying that
$d_{w} = (\ln\,5)/(\ln\,2)$ on this graph. One of the  objectives of the present
work is to examine how the effective random walk dimension is affected by
 the fact that the walker is mortal, i.e., $q <1$.

\subsection{Mortal random walker}
\label{mortalrw}

We now consider the case of a mortal random walker,
with any specified value of $q \in (0, 1)$.
Let $F_{j}(t,q)$ be the probability that a walker starting from the site $i$ hits any of the traps for the
first time at  time $t$. (This notation helps us keep track of the fact
that the first-passage time
distribution is $q$-dependent).   Since the first jump of the walker
from the site $j$ to any of the NN sites of $j$ occurs with probability
$q/\nu_{j}$, the  BKE for $F_{j}(t,q)$ is
now given by
\beq
F_{j}(t+1, q) = \frac{q}{\nu_j} \sum_{k} \delta_{\langle jk \rangle}\,F_{k}(t,q).
\label{bke3}
\eeq
The  boundary conditions and initial conditions on $F_{j}(t, q)$ are the same as those
satisfied by $\phi_{j}(t)$: namely,
$F_{j}(t,q) = \delta_{t,0}$ when $j$ is  a trap site, and
$F_{j}(0,q) = 0$ when $j$ is a non-trap site.
Equation  \eqref{bke3} differs from the BKE \eqref{bke1}
solely  by the extra factor of $q$
on the right-hand side.  As  the respective time arguments on the left
 and right-hand sides
of Eq. \eqref{bke3} are $t+1$ and $t$,
the presence of this factor implies at once
that $F_{j}(t, q)$ must necessarily
 be  of the form $q^{t}\,\phi_{j}(t)$, where $\phi_{j}(t)$ is the
 first-passage-time distribution in the case $q = 1$, as already defined.
The total probability of the trapping of a mortal walker
starting from $i$ is therefore given by
\beq
P_{j}(q) = \sum_{t = 0}^{\infty} F_{j}(t,q) = \sum_{t = 0}^{\infty} q^{t} \phi_{j} (t) < 1,
\label{unsuretrapping}
\eeq
because  $\phi_{j}(t)$ is already normalized to unity
($P_{j}(1)  \equiv \Phi_{j} = 1$),  and $0 < q < 1$.
The trapping of a walker starting from an arbitrary initial site
is therefore
no longer  a sure event,
and first passage to a trap from an arbitrary initial site is not a proper random variable.
The total probability of reaching a trap  depends on the starting point $j$.
But we can still define a mean first-passage time (MFPT) or mean
time to trapping, $T_{j}(q)$,  by averaging over the set of realizations of the walk starting from
$j$ in which trapping  does occur. This requires the first moment of $F_{j}(t,q)$
 to be divided by the total trapping probability $P_{j}(q)$.
 As  this  denominator is different for different sites,
 one can no longer expect any linear relation between the MFPTs, in general.
 Further,   simple multiplicative scaling relations will no longer hold
 for MFPTs on hierarchical graphs.
 But, as will be seen in the sequel, the self-similarity of such graphs does lead to
 more intricate scaling relations, involving in each case the iterates of a scaling function.
 It will also become clear that the  case $q = 1$ is in a separate class
 by itself, in a certain specific sense.

 It is evident from Eq. \eqref{unsuretrapping} that $P_{j}(q)$ is the generating
 function (or `partition function') for the probability distribution $\phi_{j}(t)$,
 while $q$ plays the role of a fugacity parameter.  This fact proves to be of great help
in the analysis that follows. The mean time to trapping for walks originating at $j$
is  defined as
 \beq
 T_{j}(q) = \sum_{t=0}^{\infty} t \,F_{j}(t,q)
{\Big/} \sum_{t=0}^{\infty} F_{j}(t,q).
\label{mfptmortal1}
\eeq
Using the fact that $F_{j}(t,q) = q^{t}\,\phi_{j}(t)$,  Eq. \eqref{mfptmortal1}
can be re-written as
\beq
T_{j}(q)  =   q \,\frac{d}{dq} \,\ln\,P_{j}(q).
\label{mfptmortal2}
\eeq
Other such formulas can be written down for the higher moments of the
time to trapping from any initial site $j$, in terms of the higher derivatives
of $P_{j}(q)$ with  respect  to $q$. It remains to find an equation for the set of
trapping probabilities $\{P_{j}(q)\}$.
Summing over $t$ in Eq. \eqref{bke3} yields the equation sought.
We find
\beq
P_{j}(q) = \frac{q}{\nu_j} \sum_{k} \delta_{\langle jk \rangle}\,P_{k}(q).
\label{bke4}
\eeq
While this is a trivial relation in the case $q = 1$ (with the solution
$P_{j}(1) = 1$ for every $j$), it is far from being so for $q \neq 1$.
In particular,  it  immediately precludes the possibility that
$P_{j}(q)$ could be  independent of $j$ when $q < 1$. Once again,
the `boundary condition'
$P_{k}(q) = 1$ when $k$ is any trap site makes
\eqref{bke4} an inhomogeneous set of linear equations, guaranteed to have a unique and non-trivial solution set.

\subsection{Mortal walker on a linear lattice}
\label{linearlattice}

As a simple  illustration of the effects of mortality ($q < 1$),
consider a Markovian random
walk via nearest-neighbor jumps
on a linear lattice with sites labeled $0, 1, \ldots, N$, with a trap at $N$.
The total probability that a walker starting at the site $j$ hits the trap is given by
$P_{j}(q)$. The   set of equations   \eqref{bke4} reads, in this instance,
\beq
 P_{0}(q)  = q P_{1}(s)
 \label{bkelinear1}
 \eeq
 and, for $1\leq  j \leq N-1$,
 \beq
  P_{j}(q)  =   \tfrac{1}{2} q [P_{j-1}(q) +
 P_{j+1}(q)],
 \label{bkelinear2}
\eeq
 with the  boundary condition $P_{N}(q) = 1$.
The last equation of the set is therefore
 \beq
 P_{N-1}(q) = \tfrac{1}{2} q[P_{N-2}(q) + 1],
 \label{bkelinear3}
 \eeq
which  is an inhomogeneous equation.  Hence, there is a unique non-trivial
 solution set for $\{P_{j}(q)\}$.   As we know already,
in the case $q = 1$ this is just the
uniform solution $P_{j}(1) = 1$ for every $i$, but this is no longer true for any $q < 1$. In fact,
Eq.  \eqref{bkelinear2} is precisely the recursion relation satisfied by the
Chebyshev polynomials  of the first and second kinds, with argument
$1/q$.
The conditions (i) \, $P_{N}(q) = 1$  and  (ii)\, $0 \leq P_{j}(q) < 1$
suffice to identify the unique normalized solution to be
\beq
P_{j}(q) = \frac{\mathcal{T}_{j}(1/q)}{\mathcal{T}_{N}(1/q)}\,,
\label{linearlatticesoln}
\eeq
where $\mathcal{T}_{j}(x)$ is the Chebyshev polynomial of the first kind and of order $j$.
Since $1/q  \geq 1$, we have the representation
\beq
\mathcal{T}_{j}(1/q) = \cosh\,\big(j\,\cosh^{-1}\,(1/q)\big),
\label{chebyshev1}
\eeq
showing that $P_{j}(q) < P_{j^{\prime}}(q)$ when $j < j^{\prime}$.
This is just what is  expected on
physical grounds: the trapping probability increases as the starting point of the walk gets closer to the trap. In the limit  $q = 1$, Eq. \eqref{linearlatticesoln} yields
$P_{j}(1) = 1$, as required. At the other extreme ($q\rightarrow 0$),
$P_{j}(q)$ vanishes exponentially with the distance to the trap,
like $q^{N-j}
= \exp\,[-(N-j)\,\ln\,(1/q)]$.    \\

Using the formula  of Eq. \eqref{mfptmortal2},
the mean time taken by a mortal random walker starting at the site $j$ to reach the
trap at $N$ works out to
\beq
T_{j}(q)  =  \frac{\big\{
N\,\tanh\,(N\,\tanh^{-1}\,
\sqrt{1-q^{2}}\,)
 - j\,\tanh\,(j\,\tanh^{-1}\,
\sqrt{1-q^{2}}\,)\big\}}{\sqrt{1-q^{2}}}.
\label{mfptlinearlattice}
\eeq
Again, in the limit $q = 1$, we recover the well-known result $T_{j}(1) = N^{2} - j^{2}$
(standard diffusive behavior). In the limit $q \rightarrow 0$, we have $T_{j}(q)
\rightarrow N-j$, which suggests `ballistic' motion---
the mean time taken to reach the trap is proportional to the distance to be covered.
This turns out to be a general feature that has  a straightforward
explanation, as we shall see. For the moment, it suffices to
bear in mind  that
the average involved in this MFPT is over the vanishingly small number of
realizations of the walk in which a first passage to the trap does occur.

\section{Scaling of trapping probability on a hierarchical lattice}
\label{trappingprobscaling}

\subsection{Hierarchical linear  lattice}
\label{hierarchicallinear}
We turn now to the application of the foregoing to a mortal random walker on a hierarchical
lattice.  It is helpful to
illustrate the manner in which the probability of trapping
scales on going from one generation to the next on
a family of  hierarchical graphs comprising a suitable subset of
the set of linear lattices.   By `scaling', we mean here a sequence of
renormalizations of the original survival probability $q$.  More than one
transformation of this type can be envisaged, originating from the
nesting property
of the Chebyshev polynomials,  namely,
\beq
\mathcal{T}_{rj}(x) = \mathcal{T}_{r}\big(\mathcal{T}_{j}(x)\big),
\label{chebyshevnesting}
\eeq
where $r =  0, 1, 2, \dotsc\,$. The first nontrivial
transformation  in this regard corresponds to $r = 2$,
which we now proceed to consider in a specific form.  \\

Consider the subset  $\{G_{n}\,\vert\,n = 0, 1, 2, \ldots\}$ of linear lattices,
where the sites of the $n^{\rm th}$ generation graph $G_{n}$ are labeled from $0$ to $2^{n}$.
The generation-$0$ graph $G_{0}$ comprises just two sites, $0$ and $1$. This is decorated with a site in the middle of the bond, and the length scale doubled,
to obtain the generation-$1$ graph $G_{1}$. The procedure is repeated
to obtain the family  $\{G_{n}\}$ of hierarchical graphs.   We consider, specifically,
the  probability
$P_{0}^{(n)}(q)$
that a mortal walker starting from $0$ on  $G_{n}$
hits the trap
located at the other end of the lattice, at site $2^{n}$.
  The superscript
 $(n)$ is  meant to
 keep track of the fact that the random walk occurs on $G_{n}$.
 It is important to note that  no new traps are
 added in going from one member of the hierarchy to the next.
 The distance to be covered by the walker doubles
 from one generation to the next.
 $P_{0}^{(n)}(q)$ has
  already been
 determined in the preceding section: setting
 $j = 0$ and $N = 2^{n}$ in Eqs. \eqref{linearlatticesoln} and
 \eqref{mfptlinearlattice}, we have
 \beq
P_{0}^{(n)}(q)
= \frac{1}{\mathcal{T}_{2^{n}}(q)} = {\rm sech}\,
\big(2^{n}\,{\rm  sech}^{-1}\,q\big),
\label{gnlinearsoln}
\eeq
while the  corresponding mean time to trapping
is
\beq
T_{0}^{(n)}(q)
=\frac{2^{n}\,\tanh\,\big(2^{n}\,\tanh^{-1}\,
\sqrt{1-q^{2}}\big)}{\sqrt{1-q^{2}}}\,.
\label{gnlinearmfpt}
\eeq
The exact, explicit expressions in Eqs. \eqref{gnlinearsoln} and \eqref{gnlinearmfpt}
enable us to see precisely how the probability of trapping and the corresponding mean
time to trapping vary as functions of the survival probability $q$ of a mortal random
walker on the hierarchical linear lattice. As
$q$ increases from $0$ to $1$, $P_{0}(q)$ stays close to
$0$  and rises very slowly, and then rapidly rises up to the value $1$ at $q=1$.
The MFPT $T_{0}^{(n)}(q)$, too, exhibits a similar-shaped variation, as it
rises from its lower limiting value $2^{n}$ at $q=0$ to its upper limiting value
$(2^{n})^{2} = 2^{2n}$ at $q=1$. We will return, subsequently,  to the change in the behavior
of the MFPT (and hence that of the temporal scaling factor $\mu$) for
a mortal walker.  \\

At  the moment, however,  we are interested in deducing the  foregoing solution
for $P_{0}^{(n)}(q)$ on the basis of a
 scaling argument that can be generalized to other hierarchical graphs.
 On $G_{0}$ we  have, trivially, $P_{0}^{(0)}(q) = q$ and
$T_{0}^{(0)}(q) = 1$.  Finding $P_{0}^{(1)}(q)$ for $G_{1}$
requires, in principle,
the enumeration of all walks between the sites $0$ and $1$ before the walker hits $2$ for the
first time. This is quite easy, but
it is even easier to solve
Eqs.  \eqref{bkelinear1}--\eqref{bkelinear3}
explicitly in this case.  We find
\beq
P_{0}^{(1)}(q) = \frac{q^{2}}{(2-q^{2})}.
\label{g1linear}
\eeq
Similarly,  Eqs.  \eqref{bkelinear1}--\eqref{bkelinear3}
can be solved explicitly  on $G_{2}$ and
$G_{3}$
to arrive at the solutions
\beq
P_{0}^{(2)}(q) = \frac{q^{4}}{(8- 8q^{2}+ q^{4})}
\label{g2linear}
\eeq
and
\beq
P_{0}^{(3)}(q) = \frac{q^{8}}{(128-256q^{2}+ 160q^{4}-32q^{6}+ q^{8})}.
\label{g3linear}
\eeq
The number of
equations in \eqref{bkelinear2}
increases exponentially with increasing $n$, making a brute-force solution
of the set  of equations \eqref{bkelinear1}--\eqref{bkelinear3} intractable.
But we note that, in going from $G_{0}$ to $G_{1}$,
the probability of survival of the
walker till it reaches the trap decreases,
from the value $q$ on $G_{0}$ to the value $q^{2}/(2-q^{2})$ on $G_{1}$.
In other words,
as the distance between the starting point and the trap is {\em doubled},
the survival parameter $q$ is effectively   {\em rescaled} to a new value
according to the map
\beq
q \rightarrow f(q) = \frac{q^{2}}{(2 - q^{2})}.
\label{flinear}
\eeq
We therefore expect the solutions in Eqs. \eqref{g2linear} and
\eqref{g3linear} to be the  iterates
$f(f(q))$ and $f(f(f(q)))$ of
the map $f(q)$, and  it is readily verified that this is
indeed  so. Owing to the
exact hierarchical nature of the set $\{G_{n}\}$ and of the
locations of the initial and final sites of the random walk,
the probability of
a walker on $G_{n}$ starting at the site $0$ and hitting the trap at
site $2^{n}$ should then be
given by
\beq
P_{0}^{(n)}(q) =
f\big(P_{0}^{(n-1)}(q)\big) =
f^{(n)}(q),
\label{gnlinear}
\eeq
where $f^{(n)}(q)$ is the $n^{\rm th}$ iterate of the map $f(q)$
(with $f^{(0)}(q) \equiv q$).  But this is exactly what we have already proved:
noting that  $f(q) = 1/\mathcal{T}_{2}(1/q)$,  we have
\beq
f^{(2)}(q) = f\big(1\big/\mathcal{T}_{2}(1/q)\big)
= \frac{1}{\mathcal{T}_{2}\big(\mathcal{T}_{2}(1/q)\big)}
= \frac{1}{\mathcal{T}_{4}(1/q)} = P_{0}^{(2)}(q),
\label{}
\eeq
and so on, successively.  The assertion  that $P_{0}^{(n)}(q) =
f\big(P_{0}^{(n-1)}(q)\big)$ follows from the nesting property of the
Chebyshev polynomials (Eq. \eqref{chebyshevnesting}),
i.e., from the fact that
$\mathcal{T}_{2}\big(\mathcal{T}_{2^{n-1}}(1/q)\big)
= \mathcal{T}_{2^{n}}(1/q)$.

With this `scaling solution'
 at hand, the focus shifts to the analysis of
the map $f(q)$. In  the particular example of the
linear hierarchical lattice, we already have the explicit form of $f^{(n)}(q)$
as a function of $q$ for an arbitrary value of $n$. But
such a form  is not available for an arbitrary  hierarchical lattice. It is therefore
necessary to work out a general formalism that enables deductions to be made even in the
absence of an explicit solution, as we now proceed to show.
The exact solution \eqref{gnlinearsoln}  (pertaining to the
hierarchical linear lattice)  and its properties will then serve to corroborate
the results to be deduced on general grounds.

In the unit interval $[0,1]$ in $q$
(the physical region), the map $f(q)$ in Eq. \eqref{flinear} is onto, monotone and convex,
with a superstable fixed point at $q = 0$ and an  unstable fixed point
at  $q = 1$. Thus, if $n <  m$, then $f^{(n)}(q) > f^{(m)}(q)$
for every  $0 < q < 1$. As the generation number $n$ increases,
any initial $s <1$ flows into the fixed point
at  $q = 0$.  Correspondingly, $P_{0}^{(n)}(q)$ becomes flatter and flatter over the interval, with
a leading behavior
\beq
P_{0}^{(n)}(q) \sim 2\,(q/2)^{2^{n}}
\label{linearasympprob}
\eeq
 near $q = 0$, and rises steeply as $q \rightarrow 1$
to reach the value $1$ at the fixed point $q = 1$.
This is also the leading large-$n$ behavior of $P_{0}^{(n)}(q)$ for any $q < 1$,
because of the flow toward the stable fixed point with increasing $n$.
The only exception to this behavior
corresponds, of course, to the case $q = 1$, which remains fixed at that value
under iteration. It is in this sense that this case  remains distinct from that of a mortal walker
with any value of $q$ less than unity. For a mortal walker, the probability
of reaching the trap decreases exponentially with increasing  distance from the origin,
since the distance from the origin to the trap  is $2^{n}$. The characteristic
length scale of this exponential decay is $1/\ln\,(1/q)$.

The iterative form of  $P_{0}^{(n)}(q)$ in
Eq.  \eqref{gnlinear} also leads to a useful expression for the mean time to trapping
for a random walk starting at the site $0$. From Eq. \eqref{mfptmortal2}, we have in this case
\beq
T_{0}^{(n)}(q) = q\,\frac{d}{dq}\,\ln\,f^{(n)}(q).
\label{mfptlinear1}
\eeq
Let the sequence
$q_{0}\xrightarrow[]{f} q_{1}  \xrightarrow[]{f} q_{2}
\dotsb \xrightarrow[]{f} q_{n}$ denote the
orbit of the point $q_{0}\equiv q$ under the map $f$, i.e.,
\beq
q_{\alpha} \equiv  f^{(\alpha)}(q), \;\;\alpha = 0, 1, \dotsc , n.
\label{salphadefn}
\eeq
Equation \eqref{mfptlinear1} can then be written,  for $n \geq 1$, as
\beq
T_{0}^{(n)}(q) = \frac{q_{0}}{q_{n}}\,\frac{dq_{n}}{dq_{0}}
= \frac{q_{0}}{q_{n}}\,\frac{dq_{n}}{dq_{n-1}}\,\frac{dq_{n-1}}{dq_{n-2}}
\cdots \frac{dq_{1}}{dq_{0}}\,.
\label{mfptlinear2}
\eeq
Since $dq_{\alpha+1}/dq_{\alpha} =
f^{\prime}(q_{\alpha})$
(where  the prime denotes  the derivative),
we get
\beq
T_{0}^{(n)}(q)
= \frac{q_{0}}{q_{n}}\,\prod_{\alpha=0}^{n-1} f^{\prime}(q_{\alpha}), \;\;n \geq 1.
\label{mfptlinear3}
\eeq
The time to trapping can thus be expressed in terms of
a product of the local contraction  factors pertaining to the map,
evaluated at the successive points on the orbit of
$q_{0}$.  The MFPT  $T_{0}^{(n)}(q)$ also provides us with a
natural choice for the temporal scaling factor characterizing
a mortal random walker on a family of hierarchical graphs, as we go
from one generation to the next. We define
\beq
\mu = \frac{T_{0}^{(n)}(q)}{T_{0}^{(n-1)}(q)}\,.
\label{mudefinition}
\eeq
Using the expression in  Eq. \eqref{mfptlinear3} in this definition, we get
the very convenient formula
\beq
\mu=
\Big[q\,\frac{d}{dq}\,\ln\,f(q)\Big]_{q=q_{n-1}}\,.
\label{muformula}
\eeq
The formulas in Eqs. \eqref{gnlinear},  \eqref{mfptlinear1}--\eqref{muformula}
are of general applicability to mortal random walks on hierarchical
lattices with the appropriate scaling function $f(q)$ in each case.
They provide the basis for what follows in the sequel.

Applying the formula of Eq. \eqref{mfptlinear3} to the  map \eqref{flinear}
corresponding to the hierarchical linear lattice, we have
\beq
T_{0}^{(n)}(q) = \frac{4^{n}\,q_{0}}{q_{n}}\,\prod_{\alpha=0}^{n-1}
\frac{q_{\alpha}}{(2-q_{\alpha}^{2})^{2}}\,.
\label{mfptlinear4}
\eeq
In the case $q=1$ we have $q_{\alpha}=1$ for every $\alpha$, and the standard
random walk result $T_{0}^{(n)}(1) = (2^{n})^{2}$ is recovered: the MFPT to
traverse a distance $2^{n}$ is just the square of that distance. But this is no longer
true for any $q < 1$. The respective spatial and temporal
scaling factors $\lambda$ and $\mu$ for a mortal random
walker on the family of hierarchical linear
lattices are deduced readily.  In going from $G_{n-1}$ to $G_{n}$,
the length of the lattice is simply doubled, so that $\lambda = 2$.
Using  Eq. \eqref{muformula},  the
 corresponding temporal factor is
\beq
\mu = \frac{4}{2-q_{n-1}^{2}}\,.
\label{mulinear1}
\eeq
For $q=1$, of course, we have $q_{\alpha} =1$ for every $\alpha$, so that
$\mu = 4$, and we recover the familiar result  $d_{w}= (\ln\,\mu)/(\ln\,\lambda) =
2$  (independent of $n$). But for any $q < 1$,
the ratio $\mu$ as given by Eq. \eqref{mulinear1}
 is still $n$-dependent,  in keeping with the fact that the
scaling is not a simple multiplicative one in the case of a mortal walker.
$\mu$  starts at the value $4/(2-q^{2})$ for $n = 1$, and decreases as
$n$ increases.
In the large-$n$ limit, since  any initial $q < 1$ flows toward
$q = 0$,  we find that  $\mu \rightarrow 2$.  This would imply a
 walk dimension
$d_{w} \rightarrow 1$, which is characteristic of deterministic
(ballistic) motion,
rather than diffusive motion. But there is a simple explanation for this
behavior. The leading contribution to $P_{0}^{(n)}(q)$ for large $n$,
as given by  Eq. \eqref{linearasympprob}, corresponds precisely to the
realization of a random walk from $0$ to the trap at $2^{n}$ in which the
walker never jumps back, but moves in a directed path from start to finish.
There is only one such walk.
Each of the $2^{n}$ steps occurs with a probability $\frac{1}{2}q$, except the first step
from $0$ to $1$ which occurs with a probability $q$. Hence, the probability of this
walk is $2 (q/2)^{2^{n}}$, and the time taken to execute it is equal to the number of
steps in it, namely, $2^{n}$.

A final remark concerning the family of linear hierarchical lattices:
Equation  \eqref{gnlinearsoln} is an explicit functional form
for the trapping probability $P_{0}^{(n)}(q)$.  The latter is the
$n^{\rm th}$ iterate $f^{(n)}(q)$ of the map $f(q)$. It  is
therefore clear  that
the specific recursion relation in this case, namely,
\beq
q_{\alpha+1} = f(q_{\alpha}) = \frac{q_{\alpha}^{2}}{(2 -q_{\alpha}^{2})}\,,
\label{flinearrecursion}
\eeq
must actually be solvable  in terms of elementary functions.
Setting $q_{\alpha} = {\rm sech}\,\theta_{\alpha}$, we have
$\cosh\,\theta_{\alpha +1} = 2\,\cosh^{2}\theta_{\alpha}- 1 =  \cosh\,2\theta_{\alpha}$,
so that $\theta_{\alpha} = 2^{\alpha}\,\theta_{0}$. It follows at once that
$q_{n}  =  {\rm sech}\,\big(2^{n}\,{\rm  sech}^{-1}\,q\big)$,
which is precisely Eq. \eqref{gnlinearsoln}.

\subsection{Mortal walker on the Sierpinski gasket}
\label{gasket}

\begin{figure}[!h]
\includegraphics[width=300px]{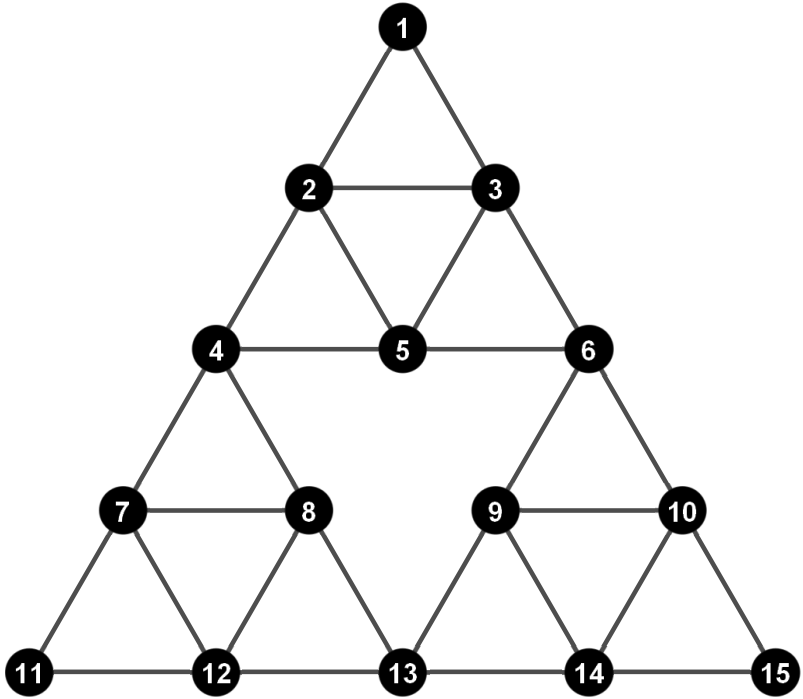}
\caption{\label{fig:first}Sierpinski Gasket $N\equiv N_2=15$}
\label{first}
\end{figure}

We turn, now, to the case of a mortal random walker on a prototypical hierarchical
lattice, the family of Sierpinski graphs embedded in $d = 2$ dimensions. The
procedure for constructing the family $\{G_{n}\}$ in this case is well known.
$G_{0}$ comprises $3$ sites
forming an equilateral  triangle with sides of unit length: the apex site $A$, and the
sites $L$ and $R$ on the base of the triangle.
$G_{1}$ is generated  by  decorating each side  with
a site at its midpoint,  joining it to its four nearest-neighbor sites with
bonds,  and doubling the  length scale.  Repeating this  process
of decorating each bond with a fresh site and doubling the
length scale generates the family of  planar Sierpinski graphs.
The  $n^{\rm th}$ generation
graph $G_{n}$ has   $N_{n}=  \frac{3}{2}(3^{n}+1)$ sites,
with $A, L$ and $R$ as the vertices of
the outermost triangle whose side length is $2^{n}$.
It  is convenient to number
the sites from $A\, (i=1)$ downwards, and from left to right in each horizontal row.
Thus  $A, L$ and $R$ correspond respectively to $i = 1,2$ and $3$
on $G_{0}$, and to $i = 1, 4$ and $6$   on $G_{1}$, and so on (see Fig. \ref{fig:first} for a representation of $G_2$).

 As already stated, our primary objective here
 is to analyze how the survival parameter  $q$  of a mortal walker is
 transformed  in going from one generation of $G_{n}$ to the next,
 leading to a solution for an appropriate
trapping probability  on $G_{n}$  in the form of the $n^{\rm th}$ iterate of
a certain scaling function.
  The simplest way to do so is to
consider the first-passage time distribution
for a walk that starts at $A$ and ends at traps located at $L$ and $R$.
Thus the configuration comprising the starting site of the random walk, the
locations of the traps, and of course the graph itself, is exactly self-similar
as we go from one generation to the next.
It remains to find the precise scaling function whose iterates yield the
trapping probability as a function of $q$.
The notation  we use parallels that  in the preceding sections.

A very brief recollection of the standard case $q = 1$ is again helpful.
Since in this case first passage to  $L$ or $R$ is a sure event for a walker starting at any $i$ on
$G_{n}$, the corresponding FPT distribution
$\phi_{i}^{(n)}(t)$  is normalized to unity  ($\Phi_{i}^{(n)} = 1$), and
we may work with the MFPTs directly, using
$\Delta_{ij}\,T_{j}^{(n)} = -1$ (Eq. \eqref{usualmfptbke}).
On $G_{0}$, we have $T_{A}^{(0)}
\equiv T_{1}^{(0)} = \frac{1}{2} + \frac{1}{2} =1$,
since $L$ and $R$ are traps. On $G_{1}$,
\eqref{usualmfptbke} is a set of three equations for
$T_{i}^{(1)}, \; i = 1,2,5$ (since
$T_{2}^{(1)} =  T_{3}^{(1)}$ by an obvious symmetry).
These equations are easily solved to give
$T_{A}^{(1)} \equiv T_{1}^{(1)}  = 5$.
Owing to the exact
self-similarity of the configuration,  this suffices to
enable the assertion that
$T_{A}^{(n)}= 5^{n}$. Hence $\mu = 5$, while $\lambda = 2$,
yielding the well-known result  that the random walk dimension is
$d_{w}= (\ln\,5)/(\ln\,2)$ for the family of Sierpinski gaskets in $d = 2$.

We turn now to  the case of a mortal random walker on $\{G_{n}\}$.
We  must now first compute  the total probability
$P_{A}^{(n)}(q) \equiv P_{1}^{(n)}(q)$  that a walker starting at $A$  hits one of  the traps at
$L$ and $R$,  using the corresponding BKE, Eq. \eqref{bke4}.
On $G_{0}$, this is given by $P_{A}^{(0)}(q) = \frac{1}{2}q  +\frac{1}{2}q  = q$.
On $G_{1}$ we have, omitting (for notational simplicity)
the supercripts and $q$-dependence for a moment,
and using the obvious symmetry $P_{2} = P_{3}$,
the relations
$P_{1} = qP_{2}, \; P_{2} = \frac{1}{4}q (P_{1}+ P_{2}+ P_{5}+1),\;
P_{5}= \frac{1}{2} q(P_{2}+ 1)$.
Solving for $P_{1}$, we have
\beq
P_{A}^{(1)}(q) = \frac{q^{2}}{(4-3q)}\,.
\label{g1sg}
\eeq
Going on to solve the corresponding equations on $G_{2}$ and $G_{3}$ we find,
after some algebra,
\beq
P_{A}^{(2)}(q) = \frac{q^{4}}{(4-3q)(16-12q-3q^{2})}
\label{g2sg}
\eeq
and
\begin{align}
P_{A}^{(3)}(q)& = \;q^{8}\big/\big[(64-96q+24q^{2}+9q^{3})\times\nonumber\\[4pt]
 & \times  (256-384q+96q^{2} + 36q^{3}- 3q^{4})\big].
\label{g3sg}
\end{align}
Equations \eqref{g1sg}-\eqref{g3sg} are the analogs of Eqs. \eqref{g1linear}-\eqref{g3linear}
derived earlier for the hierarchical linear lattice. Once again, we note
that these expressions are, respectively,  precisely the
iterates $f^{(2)}(q)$ and $f^{(3)}(q)$ of the map
\beq
q\rightarrow f(q) = \frac{q^{2}}{(4-3q)}\,,
\label{fsg}
\eeq
as we may anticipate from the results derived in the case of the hierarchical linear
lattice. The probability that a mortal walker starting from $A$ on the $n^{\rm th}$ generation
Sierpinski gasket eventually hits  one of the traps at the vertices $L$ and $R$ is given by
\beq
P_{A}^{(n)}(q) = f^{(n)}(q),
\label{gnsg}
\eeq
where $f^{(n)}$ stands for the $n^{\rm th}$ iterate of the map $f$ in Eq.  \eqref{fsg}.

Solving the recursion relation
\beq
q_{\alpha+1}= f(q_{\alpha}) = \frac{q_{\alpha}^{2}}{(4-3q_{\alpha})}
\label{sgrecursion}
\eeq
does not seem possible,  as opposed to the case of the recursion relation \eqref{flinearrecursion} for the
hierarchical linear lattice, which allowed one to obtain  $q_{\alpha}$ explicitly as a function of $q_{0}$ in terms of
elementary functions.  In spite of this, a good deal can be said about the behavior of the
iterates of $f$ for large generation number $n$.
In the unit interval $[0,1]$ of $q$, the map $f(q) = q^{2}/(4-3q)$
has essentially the same qualitative behavior as the map
$f(q) = q^{2}/(2-q^{2})$ characterizing the hierarchical linear lattice. Once again, we have
an onto, monotone, convex map with a superstable attractor  at $q = 0$
(because $f^{\prime}(0) = 0$)  and a repellor
at  $q = 1$. Any initial value $q_{0} = q < 1$ flows into $0$
with increasing generation number $n$. Since the Taylor expansion of $f(q)$ about
$q=0$ starts with a term that is of order $q^{2}$,
B\"ottcher's Theorem guarantees the existence  of a
function $\psi(q)$ that is analytic in a neighborhood of $q=0$,
vanishes at $q = 0$, and satisfies the functional equation
\beq
\psi\big(f(q)\big) =
\psi\big(q^{2}/(4-3q)\big) =
\big(\psi(q)\big)^{2}.
\label{sgbottchereqn}
\eeq
It follows immediately that $P_{A}^{(n)}(q) =f^{(n)}(q)\equiv q_{n}$ is of the
form
\beq
P_{A}^{(n)}(q) = \psi^{-1}\big[\big(\psi(q)\big)^{2^{n}}\big],
\label{sgformalsoln}
\eeq
where $\psi^{-1}\big(\psi(q)\big) \equiv   q$.
As before, $P_{A}^{(n)}(q)$ rises very slowly from $0$  with increasing $q$, and then
rapidly increases to unity as $q$ approaches $1$ from below. Its asymptotic
behavior near $q = 0$, and equivalently its leading large-$n$ behavior for any $q < 1$,
may be deduced from Eqs. \eqref{sgbottchereqn} and \eqref{sgformalsoln}. We find,
in the neighborhood of $q=0$,
\beq
\psi(q) = \tfrac{1}{4} q + \tfrac{3}{32} q^{2} +
\mathcal{O}(q^{3}).
\label{psisg}
\eeq
The corresponding inverse function is
\beq
\psi^{-1}(q) = 4q - 6q^{2} + \mathcal{O}(q^{3}) .
\label{inversepsisg}
\eeq
The trapping probability on $G_{n}$ is then given by
\beq
P_{A}^{(n)}(q)
=  4 \left(\frac{q}{4}\right)^{2^{n}}\big[1 + (3\times 2^{n-3})\,q + \mathcal{O}(q^{2})\big]\,.
\label{sgasympprob}
\eeq
The decay of the trapping probability $P_{A}^{(n)}(q)$ with
the distance to the traps
is again exponential in the distance, with a characteristic length scale
$1/(\ln\,q^{-1})$.  The case $q = 1$ is an exception, of course,
as it is a fixed point of the map $f(q)$.

The mean time to trapping (at $L$ or $R$) of a mortal random walker starting at
the apex $A$ of the Sierpinski graph $G_{n}$ can also be evaluated, since
the  general formula in Eq. \eqref{mfptlinear3}
is immediately applicable to the case at hand, with
$f(q) = q^{2}/(4-3q)$. We obtain the expression
\beq
T_{A}^{(n)}(q)
 = \frac{q_{0}}{q_{n}}\prod_{\alpha = 0}^{n-1}\frac{q_{\alpha}\,(8-3q_{\alpha})}
{(4-3q_{\alpha})^{2}}\,,
\label{mfptsg}
\eeq
where  $q_{\alpha} = f^{(\alpha)}(q)$.
The formula \eqref{muformula} yields, for
the  temporal scaling factor  $\mu$ for a mortal walker
on the Sierpinski graph,
\beq
\mu = \frac{T_{A}^{(n)}(q)}{T_{A}^{(n-1)}(q)}
=   \frac{8-3q_{n-1}}{4-3q_{n-1}}\,.
\label{musg1}
\eeq
It follows at once that, in the standard case $q = 1$ (in which every $q_{\alpha}= 1$),
we recover the value $\mu = 5$, and hence
 the customary result $d_{w} = (\ln\,5)/(\ln\,2)$ for the Sierpinski gasket.
On the other hand, for any mortal walker ($q<1$), the temporal scaling factor
depends on the generation number $n$ as well as on the single-step survival probability $q$.
It starts at the value $(8-3q)/(4-3q)$ for $n = 1$, and decreases as $n$ increases.
Once again, any initial $q< 1$ flows into the attractor at $q = 0$
in the large-$n$ limit, we see that $\mu \rightarrow 2$, and hence $d_{w}
\rightarrow 1$ in this regime.  The explanation, as in the preceding instance, lies in the
leading behavior of $P_{A}^{(n)}(q)$ for large $n$: this probability is dominated by that of a
random walk in which the walker starts at $A$ and proceeds in a straight line
along the sites on the outermost triangle of $G_{n}$, without jumping back
or moving to any internal site on the graph, till the walker reaches either $L$ or $R$.
There are only $2$ such walks, from $A$ to $L$ and from $A$ to $R$, respectively.
On either of them, the probability of the first step out of $A$ is $\frac{1}{2}q$, while the
probability of each of the remaining $2^{n}-1$ steps is $\frac{1}{4}q$. Hence the total
probability of this pair of paths is $2\times (q/2) \times (q/4)^{2^{n}-1} =
4(q/4)^{2^{n}}$, as in Eq. \eqref{sgasympprob}. The length of each path
($= 2^{n}$) is equal to the number of
time steps taken to traverse it, which is why $d_{w}$ formally tends to unity
in this limit.

\subsection{Mortal walker on the Sierpinski tower}
\label{tower}

\begin{figure}[!h]
\includegraphics[width=350px]{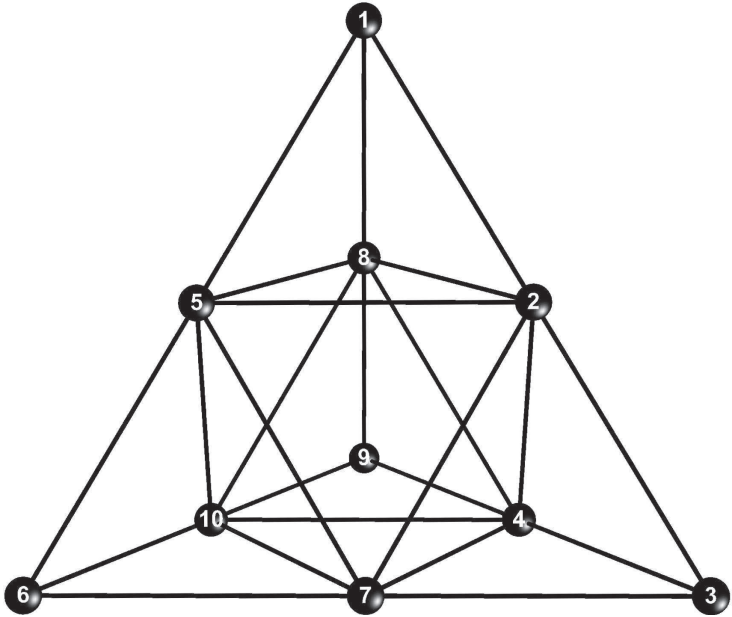}
\caption{\label{second}Sierpinski Tower $N=10$}
\label{fig:second}
\end{figure}

 It is interesting, from the theoretical point of view as
 well as that of applications, to extend the analysis in the foregoing to a
 fractal graph embedded in $d = 3$ dimensions. The natural choice is the so-called
 Sierpinski tower, constructed in a hierarchical manner similar to that
 used for the Sierpinski gasket in $d = 2$.  We begin with  $G_{0}$,  a
 tetrahedron of unit side length, its vertices being labeled $A$ (the apex)
 and $B, C, D$ (the vertices on the basal triangle). Each bond is then decorated with
 a site at its mid-point, all nearest-neighbor sites joined by bonds, and the
 length scale doubled, to obtain $G_{1}$ (see Fig. \ref{fig:second}). Iteration of this process
 yields the hierarchical family $\{G_{n}\}$ of Sierpinski towers. $G_{n}$ has
 $2(4^{n}+1)$ vertices, each with $6$ nearest neighbors, except for the
 outermost vertices $A, B, C$ and $D$, which have $4$ nearest neighbors each.
The length scale factor connecting successive generations of the family of graphs is of course
$\lambda = 2$. The corresponding temporal scale factor is known to be $6$, so that
the random walk dimension is $d_{w}= (\ln\,6)/(\ln\,2)$ for the family of Sierpinski towers.

The self-similar
configuration (comprising  the initial position of a mortal random walker on $G_{n}$ and
the locations of the traps) that is the counterpart
 of that considered in the case of the gasket is as follows.
The walker starts from $A$, and the traps are located at $B, C$ and $D$. We seek the
total probability $P_{A}^{(n)}(q)$ that a mortal walker gets trapped.
On $G_{0}$, we have $P_{A}^{(0)}(q) = 4\times \frac{1}{4}q = q$
(and hence the MFTP $T_{A}^{(0)}(q) = 1$). To find
$P_{A}^{(1)}(q)$, we must write down the BKE \eqref{bke4} for each of
the $7$ sites of $G_{1}$
(including $A$) other than the trap sites, together with the `boundary' conditions
$P_{i}^{(1)}(q) = 1$ for $i = B, C, D$.  Solving these coupled equations, we find
\beq
P_{A}^{(1)}(q) =
\frac{q^{2}}{6 - 6q + q^{2}}\,.
\label{g1st}
\eeq
Hence $P_{A}^{(1)}(1) = 1$, as expected: when $q = 1$, trapping (or first passage from $A$ to one of the
traps) is a sure event.

As in the preceding instances,  we may conclude that the survival
probability parameter $q$ is effectively renormalized as we go from
$G_{0}$ to $G_{1}$ according to the map
\beq
q \rightarrow f(q) = \frac{q^{2}}{6 - 6q + q^{2}}\,.
\label{fst}
\eeq
Once again, we note that the map is onto, monotone and convex in the unit interval
$0 \leq q \leq1$, with an unstable fixed point at $q = 1$ and a superstable attractor
at $q = 0$. The total probability of the trapping of a walker starting from $A$ at one of the
traps $B, C$ and $D$ is given by
\beq
P_{A}^{(n)}(q) = f^{(n)}(q),
\label{gnst}
\eeq
the $n^{\rm th}$ iterate of the map $f(q)$. The qualitative properties of this solution are
similar to those of the corresponding solution for the Sierpinski gasket: with increasing
$n$, the iterate $f^{(n)}(q)$ stays extremely close to $0$ for most of the unit interval
in $q$, and rises very sharply to the value $1$ at $q = 1$.  While the recursion relation
\beq
q_{\alpha+1}= \frac{q_{\alpha}^{2}}{(6 - 6q_{\alpha}+ q_{\alpha}^{2})}
\label{strecursion}
\eeq
cannot be solved to find $q_{\alpha}$ explicitly as a function of $q_{0}$, we are guaranteed
that there exists a formal solution
\beq
P_{A}^{(n)}(q) = \psi^{-1}\big[\big(\psi(q)\big)^{2^{n}}\big],
\label{stformalsoln}
\eeq
where $\psi(q)$ satisfies the functional equation
\beq
\psi\big(q^{2}/(6-6q+q^{2})\big) =
\big(\psi(q)\big)^{2}.
\label{stbottchereqn}
\eeq
$\psi(q)$ is analytic in the neighborhood of $q=0$, and has the
Taylor expansion
\beq
\psi(q) = \tfrac{1}{6} q + \tfrac{1}{12} q^{2} + \mathcal{O}(q^{3})
\label{psist}
\eeq
in that neighborhood. Its inverse function is
\beq
\psi^{-1}(q) = 6q -18 q^{2}+ \mathcal{O}(q^{3}).
\label{inversepsist}
\eeq
From Eq. \eqref{stformalsoln}, it follows that
the trapping probability on $G_{n}$ is given by
\beq
P_{A}^{(n)}(q)
= 6 \Big(\frac{q}{6}\Big)^{2^{n}} \,\big[1 + 2^{n-1}q + \mathcal{O}(q^{2})\big].
\label{stasympprob}
\eeq
Analogous to the case of the gasket, the leading large-$n$ asymptotic
contribution to $P_{A}^{(n)}(q)$ comes from
the four directed walks from $A$ to $B, C$ and $D$, respectively.
The number of steps in each of these
is $2^{n}$, and the corresponding total probability is  $4\times (q/4)\times (q/6)^{2^{n}-1}
= 6\,(q/6)^{2^{n}}$.

The formula of Eq. \eqref{mfptlinear3} gives, for the mean time to trapping
for a mortal walker starting from $A$ on $G_{n}$,  the expression
\beq
T_{A}^{(n)}(q) =
\frac{6^{n}\,q_{0}}{q_{n}}\,\prod_{\alpha=0}^{n-1}
\frac{q_{\alpha} (2-q_{\alpha})}{(6-6q_{\alpha} + q_{\alpha}^{2})^{2}}\,.
\label{mfptst}
\eeq
Using  Eq. \eqref{muformula},
the temporal scaling factor on the Sierpinski tower is
found to be
\beq
\mu
= \frac{T_{A}^{(n)}(q)}{T_{A}^{(n-1)}(q)}
= \frac{6(2-q_{n-1})}{6 - 6q_{n-1} + q_{n-1}^{2}}\,.
\label{must}
\eeq
When $q = 1$,  $\mu$ has the value $6$ independent of the generation number
$n$ (and hence $d_{w} = (\ln\,6)/(\ln\,2)$ for the
Sierpinski tower in $d = 3$). For a mortal walker, $\mu$ starts at the value
$6(2-q)/(6-6q+q^{2})$ for $n = 1$, and  approaches its limiting (`ballistic') value
$2$ as $n$ increases, for the same reason as in the preceding instances.

\section{Unconditional walk length of a mortal walk on the Sierpinski gasket}

\subsection{Preliminary remarks}

The linear system \eqref{bke4} allows one to evaluate the set $\left\{P_j\right\}$ for different starting sites. In dealing with the SG, the focus of attention so far have been the quantities $P_A^{(n)}(q)$ and the associated MFPTs (or walk lengths) $T_A^{(n)}(q)$, which refer to a setting where two deep traps are each of the bottom vertices $L$ and $R$ (we recall that the superscripts indicate that the former quantities refer to the $n$-th generation gasket). As soon as $q<1$, the walk length $T_A^{(n)}(q)$ is conditional on the walker's surviving, so that it is able to reach either of deep traps.

However, it is also of interest to consider an even simpler situation where \emph{a single} deep trap is placed at the apex site $A$ and one lets the mortal walker evolve from any site $j\neq A$ until it is \emph{either} trapped at $A$ \emph{or} it dies as a result of the mortality constraint. The set of walk lengths $\left\{T_j^{(n)}(q)\right\}$ is then of interest, and in particular the walk length $T_L^{(n)}(q)$ referring to the left bottom vertex $L$, say, as initial condition [which by symmetry is identical with $T_R^{(n)}(q)$]. A possible physical situation that corresponds to the above setting involves a perfect detector of radioactive particles placed at site $A$. Assuming that the detector clock is set to zero at the instant where the particle starts diffusing at $L$, the average over the measurement times associated with each detection event will be $T_L^{(n)}$; one assumes hereby that the diffusing particle decays according to the typical exponential law associated with Eq. \eqref{diffusioneqn}.

Returning to the general case of an arbitrary graph, in the single trap setting considered above, one may want to compute the probability of absorption $P_j(q)$ at the deep trap and the conditional mean walk length $T_j$. However, one may also be interested in the \emph{unconditional} mean walk length $\widehat{T}_j$ of the random walk (measured in number of steps); that is, the length of the walk until it is terminated \emph{either} by mortality \emph{or} by absorption at $A$, no matter what happens first. In this latter case, two competing decay channels are at play. In the example of the preceding paragraph, a diffusing radioactive particle would either \emph{die} by detector trapping or by spontaneous decay into another species, and the unconditional mean walk length $\widehat{T}_j$ would then play the role of the mean particle lifetime (measured in time steps of the random walk). Obviously, in the limit $q\to 1$, the conditional and the unconditional walk lengths become identical, $\widehat{T}_j(q=1)=T_j(q=1)$.

For a P\'olya mortal walker on any finite $N$ site graph with a set of deep traps, the site-specific unconditional walk lengths $\widehat{T}_j$ are related to one another via the following set of equations [cf. Eq. (14) in Ref. \cite{AK15}]:
\beq
\widehat{T}_j=1+\frac{q}{\nu_j}\sum_k \delta_{\langle jk\rangle} \widehat{T}_k,
\label{linsysTj}
\eeq
where one takes $\widehat{T}_k=0$ if $k$ happens to be a site with a deep trap). Note that, in the limit $q\to 0$, Eqs. \eqref{linsysTj} yield $\widehat{T_j}=1$, as a result of the "1" on the right hand side. In this case, the walker jumps, and the walk continues subject to survival probability $q$.

The linear system \eqref{linsysTj} can be solved directly for the walk lengths, thus allowing one to obtain specific numerical values when the set of transition probabilities are specified. On the other hand, in Ref. \cite{AK15} a generating function method and probability conservation arguments were invoked to show that, in the case of a single deep trap, the walk lengths and the probability of absorption at the deep trap are directly related to one another via the equation
\beq
\widehat{T}_j(q)=\frac{1}{1-q}\left[1-P_j(q)\right].
\label{TandP0}
\eeq
Note that insertion of Eq. \eqref{TandP0} into \eqref{linsysTj} allows one to recover Eq. \eqref{bke4}.
Note also that a linear relation similar to \eqref{TandP0} holds for the global averages $\widehat{T}\equiv 1/(N-1)\sum_{j\neq A} \widehat{T}_j$ and $P\equiv 1/(N-1)\sum_{j\neq A} P_j$, i.e.,
\beq
\label{TandP}
\widehat{T}(q)=\frac{1}{1-q}\left[1-P(q)\right]
\eeq
While this may seem a somewhat trivial statement, we note that in the case of conditional MFPTs studied above ($q<1$), there is not such a simple relation between $P(q)$ and $T(q)$ because of the nonlinear relation between both quantities, and knowledge of $P(q)$ alone does not suffice to evaluate $T(q)$.

In the limit $q\to 1$, one has $P_j(q)$, and trapping at the deep trap is a sure event. For the particular case of the SG, the overall conditional and unconditional walk length for each gasket generation converge to the values computed in Ref. \cite{KB02PRE}, i.e.,
\beq
T^{(n)} =\frac{1}{N_n-1} \sum_{j\neq A} T^{(n)}_j =\frac{3^n 5^{n+1}+4 (5)^n-3^n}{3^{n+1}+1}.
\eeq
Our subsequent aim will be to extend the above result to the case $q<1$ by computing the corresponding global averages $\widehat{T}^{(n)}$ and $P^{(n)}$ .

\subsection{Main results}

In III.C we were able to find a scaling function describing the behavior of  $P_A^{(n)}(q)$ for the case where two deep traps were placed at the corner sites $L$ and $R$.  In the present case where the deep trap is placed at the apex site $A$ ($P_A^{(n)}(q)=1$), it is also possible to find a scaling function describing the behavior of $P_{L,R}^{(n)}(q)$.
For the zero-th generation gasket (a triangle ALR), Eqs.  \eqref{bke4} take the form
\begin{align}
P_L^{(0)}&=\frac{q}{2} P_A^{(0)}+\frac{q}{2}P_R^{(0)}, \nonumber\\
P_R^{(0)}&=\frac{q}{2} P_A^{(0)}+\frac{q}{2}P_L^{(0)},
\end{align}
subject to the aforementioned boundary condition $P_A^{(0)}=1$. The solution is $P_L^{(0)}(q)=P_R^{(0)}(q)=q/(2-q)$. For higher gasket generations one easily finds by inspection
\beq
P_L^{(n)}(q)=h^{(n)}(P_L^{(0)}), \qquad n=1,2,3,\ldots,
\eeq
where $h^{(n)}(\cdot)$ is the $n$-th iterate of the map
\beq
h(x)\equiv h^{(1)}(x)=\frac{x^2}{2+x-2x^2}.
\eeq
Further, we also define $h^{(0)}(x)\equiv x /(2-x)$, implying that $h^{(0)}(q)=P_L^{(0)}(q)$. For convenience, in what follows we shall introduce the simplified notation $h_n \equiv h^{(n)}(q)\equiv P_L^{(n)}(q), n=0,1,2,\ldots$.
As pointed out in Ref. \cite{KB02PRE}, in the case $q=1$, one has
\beq
\label{TLs0}
T_L^{(n)}=5 T_L^{(n-1)}=5^n T_L^{(0)} = 2 \times 5^{n}.
\eeq
On the other hand, one also has the crucial relation \cite{KB02PRE}
\beq
\label{lactri1}
T_{i_r}^{(n)}+T_{j_r}^{(n)}+T_{k_r}^{(n)}=T_{I_r}^{(n)}+T_{J_r}^{(n)}+T_{K_r}^{(n)}+6 \times 5^{r-1}, \qquad r=1,2,\ldots
\eeq
In terms of $T_L^{(n)}$ Eq. (lactri1) can be rewritten as
\beq
\label{lactri2}
T_{i_r}^{(n)}+T_{j_r}^{(n)}+T_{k_r}^{(n)}=T_{I_r}^{(n)}+T_{J_r}^{(n)}+T_{K_r}^{(n)}+3T_L^{(r-1)},\qquad r=1,2,\ldots
\eeq
In Eqs. \eqref{lactri1} and \eqref{lactri2}, we recall that $(i_r,j_r,k_r)$ and $(I_r,J_r,K_r)$ respectively denote the sites demarcating lacunary triangles of ascending size $r$ and the vertex sites of the triangle with $(I_r,J_r,K_r)$ as its central lacunary region.
In the case $q<1$, the analog of Eq.  \eqref{TLs0} is
\beq
\widehat{T}_L^{(n)}=\frac{1}{1-q}[1-h_n(q)].
\eeq
For  all  $q$, one finds, for example, for the $n=1$ Sierpinski gasket ($N=N_1=6$),
\begin{subequations}
\begin{align}
\widehat{T}_2^{(1)} + \widehat{T}_3^{(1)} + \widehat{T}_5^{(1)} &= \frac{3 q^2  - 16q + 24}{3 q^2 -10 q + 8}, \\
\widehat{T}_1^{(1)} + \widehat{T}_4^{(1)} + \widehat{T}_6^{(1)} &= \frac{3 q -8}{3 q - 4}.
\end{align}
\end{subequations}
Hence,
\begin{align}
\label{trianglerel}
 (\widehat{T}_1^{(1)} + \widehat{T}_4^{(1)} + \widehat{T}_6^{(1)}) - (\widehat{T}_2^{(1)} + \widehat{T}_3^{(1)} + \widehat{T}_5^{(1)}) =  \frac{2 (q-4)}{(3 q - 4) (q - 2)}.
\end{align}
When $q=1$, and for this case alone, this expression reduces to the result, Eq. (5) in Ref. (31),
\begin{align}
T_2^{(1)} + T_3^{(1)} + T_5^{(1)} = T_1^{(1)} + T_4^{(1)} + T_6^{(1)} + 6.
\end{align}
For the $n=3$ Sierpinski gasket ($N=42$),  the sum $T_2^{(3)} + T_3^{(3)} + T_5^{(3)}$ reads
\begin{align}
\frac{\left(1118208 q^3+2408448 q^2-2129920 q+688128\right) \left(12600 q^6+72192 q^5+93312
	q^4\right) \left(27 q^8-1008 q^7\right)}{108 (q-2) q^7 (3 q-4) \left(3 q^2+12 q-16\right)
	\left(96 q^2+384 q-256\right)},
\end{align}
whereas the sum $T_1^{(3)} + T_4^{(3)} + T_6^{(3)}$ is
\begin{align}
\frac{(3 q-8) \left(9 q^6-288 q^5-5664 q^4-1152 q^3+58368 q^2-92160 q+40960\right)}{(3 q-4) \left(3 q^2+12 q-16\right) \left(3 q^4-36 q^3-96 q^2+384
	q-256\right)}.
\end{align}
Thus, the factor on the right-hand side of the analog of Eq. \eqref{trianglerel} is, once again,
\begin{align}
 \frac{2(q-4)}{(3q-4)(q-2)}.
\end{align}
In more general terms, the counterpart of Eq. \eqref{lactri2} is found to be
\beq
\label{lactri3}
\widehat{T}_{i_r}^{(n)}+\widehat{T}_{j_r}^{(n)}+\widehat{T}_{k_r}^{(n)}=P_L^{(r-1)}[\widehat{T}_{I_r}^{(n)}+\widehat{T}_{J_r}^{(n)}+\widehat{T}_{K_r}^{(n)}]+3\widehat{T}_L^{(r-1)},
\qquad r=1,2,\ldots
\eeq
Or, in terms of the $h_n$'s,
\beq
\label{lactri4}
\widehat{T}_{i_r}^{(n)}+\widehat{T}_{j_r}^{(n)}+\widehat{T}_{k_r}^{(n)}=h_{r-1}[\widehat{T}_{I_r}^{(n)}+\widehat{T}_{J_r}^{(n)}+\widehat{T}_{K_r}^{(n)}]+3\frac{1-h_{r-1}}{1-q},
\qquad r=1,2,\ldots
\eeq
One is now tempted to compute the global average $\widehat{T}^{(n)}$ by methods similar to those employed in
\cite{KB02PRE}. However, at this stage we realize that the absorption probabilities $P_j^{(n)}$ referring to the sets of sites $(i_r,j_r,k_r)$ and $(I_r,J_r,K_r)$ fulfil a simpler relation, i.e.,
\beq
\label{lactri5}
P_{i_r}^{(n)}+P_{j_r}^{(n)}+P_{k_r}^{(n)}=h_{r-1}[P_{I_r}^{(n)}+P_{J_r}^{(n)}+P_{K_r}^{(n)}],
\qquad r=1,2,\ldots
\eeq
This prompts us to work with the above hierarchical relation rather than with the set of Eqs. \eqref{lactri4}. Using Eq. \eqref{lactri5} and the equivalence $P_L^{(n)}=P_R^{(n)}$, it is possible to compute the global average $P^{(n)}=(N_n-1)^{-1} \sum_{j\neq A} P_j^{(n)}$ by suitably reexpressing the site-specific probabilities of absorption at $A$ in terms of $P_L^{(n)}$ only (see Ref. \cite{KB02PRE}). Thus, to obtain the $P^{(n)}$'s, one does not need to compute site-specific probabilities other than the set $\left\{P_L^{(m)}\right\}\equiv \left\{h_m\right\}$ (with $m=0,1,\ldots,n$) , whence the global unconditional walk length $\widehat{T}^{(n)}$ immediately follows via Eq. \eqref{TandP}. For the first few generations one obtains
\begin{subequations}
\begin{align}
&n=0, \qquad P^{(0)}=h_0=\frac{q}{2-q}, \\
&n=1, \qquad P^{(1)}=\frac{(1+h_0)(1+2h_1)-1}{5}=\tfrac{4q+q^2}{40-30q-5q^2}, \\
&n=2, \qquad P^{(2)}=\frac{(1+h_0+h_1+2h_0h_1)(1+2h_2)-1}{14}=\tfrac{32 q - 24 q^2 - 2 q^3 + q^4}
{896 - 1344 q + 336 q^2 + 126 q^3 - 7 q^4}, \\
&n=3, \qquad P^{(3)}=\frac{(1+h_0+h_1+2h_0h_1+2h_0h_2+2h_1h_2+4h_0h_1h_2)(1+2h_3)-1}{41} \nonumber \\
& =\tfrac{16384 q - 36864 q^2 + 23552 q^3 - 512 q^4 - 2560 q^5 + 40 q^7 + q^8}
{1343488 - 4030464 q + 4030464 q^2 - 1133568 q^3 - 393600 q^4 +
 165312 q^5 + 20664 q^6 - 2214 q^7 - 41 q^8}.
\end{align}
\end{subequations}
For arbitrary generation number $n$, the general expression of the probability of absorption (averaged over all the sites other than the deep trap) turns out to be
\beq
\resizebox{462pt}{!}{%
$
\hspace{0.1cm} P^{({n})}=\frac{\left(1+2%
\sum^\prime_i h_i+4\sum^\prime_{\stackrel{\left\{i,j\right\}}{i\neq j}} h_{i} h_{j}%
+8\sum^\prime_{\stackrel{\left\{i,j,k\right\}}{i\neq j \neq k}} h_i h_j h_k+\ldots%
+2^{n-1} h_0h_1h_2\ldots h_{n-2}h_{n-1}\right)\left(1+2h_n\right)-1}{N_n-1},%
$
}
\eeq

where the primes in the sums indicate that the different indices $i,j,k,\ldots$ take values from
$0$ to $n-1$. Using Eq. \eqref{TandP}, one can subsequently find the unconditional walk length for the $n$-th generation gasket, i.e.,
\beq
\widehat{T}^{(n)}(q)=\frac{1}{1-q}\left[1-P^{(n)}(q)\right].
\eeq

\section{Comparison between results for conditional and unconditional walk lengths}
Both conditional and unconditional walk lengths are relevant for target search problems, hence we will compare these two quantities in the present section. Eqs. \eqref{mfptmortal2} and \eqref{bke4} are the basis to compute the site-specific conditional walk lengths, whereas Eqs. \eqref{linsysTj} can be used to compute site-specific unconditional walk lengths. We recall that the validity of Eqs. \eqref{linsysTj} implies that the walker jumps, and then the walk continues subject to survival probability $q$. This convention has been used throughout Sec. IV for the sake of comparison of the obtained unconditional walk lengths with previous results for the Sierpinski gasket in the $q\to 1$ limit. In the $q\to 0$ limit, it implies $\widehat{T}_j\to 1$, i.e., the walker jumps at least once.

However, for a comparison between conditional and unconditional walk lengths, it appears more natural to first check whether the walker has survived and, if so, then the jump is implemented. In practical terms, this means that, already before taking the first step, the walker has a non-zero probability of dying $1-q$. Consequently, the walk length becomes smaller by a factor $q$. With this new convention, Eqs. \eqref{TandP0} and Eqs. \eqref{TandP} must now be replaced with
\beq
\label{TandP0new}
\widehat{T}_j(q)=\frac{q}{1-q}\left[1-P_j(q)\right]
\eeq
and
\beq
\label{TandPnew}
\widehat{T}(q)=\frac{q}{1-q}\left[1-P(q)\right],
\eeq
respectively. On the other hand, Eq. \eqref{bke4} \emph{must} remain valid regardless of the convention used to compute the unconditional walk length, namely, the convention used here, or the one used in Sec. IV. Together with \eqref{TandP0new}, this requirement results a new equation for the $\widehat{T}_j$'s, namely,
\beq
\widehat{T}_j=q+\frac{q}{\nu_j}\sum_k \delta_{\langle jk\rangle} \widehat{T}_k,
\label{linsysTj2}
\eeq
instead, implying that all $\widehat{T_j}$ go to $0$ as $q\to 0$. Note that the only formal difference between Eqs. \eqref{linsysTj} and
\eqref{linsysTj2} is that the "$+1$" on the right hand side is replaced with "$+q$" in the latter. The unconditional walk lengths computed in the present section stem from the solution of Eqs. \eqref{linsysTj2}, but the conclusions of this section remain qualitatively the same regardless of which convention is used for the number of time steps.

\begin{figure}[!h]
	\includegraphics[width=320px]{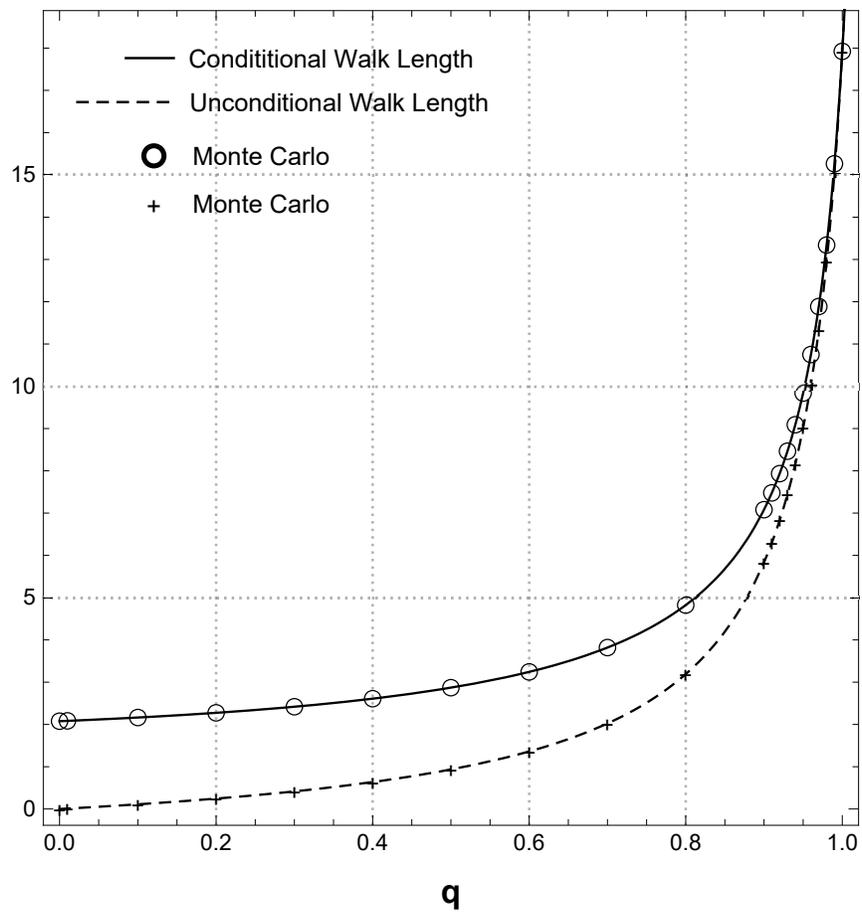}
	\caption{Mean walk length versus survival probability $q$ for the $N=15$ Sierpinski gasket (traps at two corner sites).}
	\label{fig:sierp15}
\end{figure}

\begin{figure}[!htbp]
	\includegraphics[width=320px]{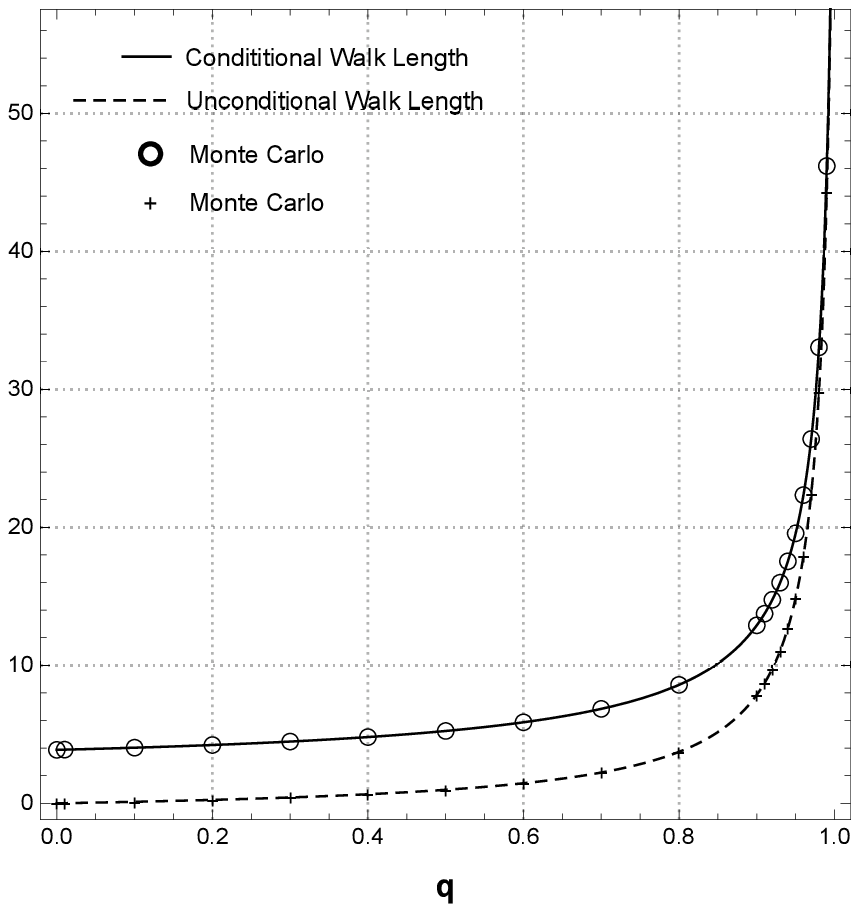}
	\caption{Mean walk length versus survival probability  $q$  for the $N=42$ Sierpinski gasket (traps at
		two corner sites)}
	\label{fig:sierp42}
\end{figure}

\begin{figure}[!htbp]
	\includegraphics[width=320px]{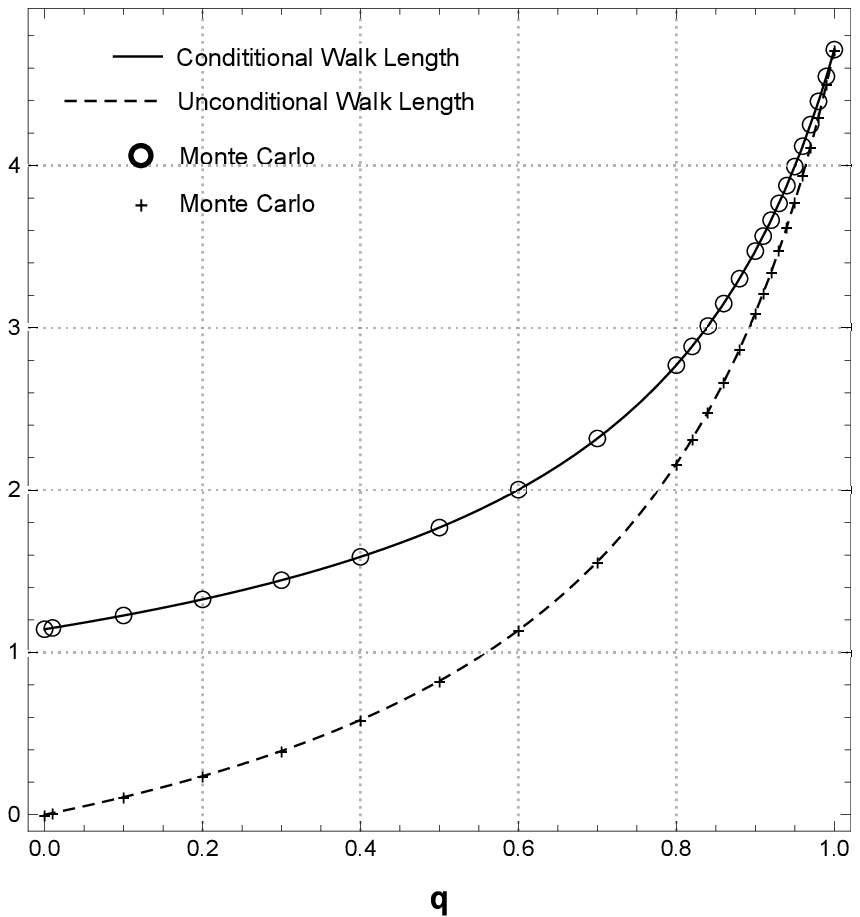}
	\caption{Mean walk length versus survival probability  $q$  for the $N=10$ Sierpinski tower (traps at
		three corner sites)}
	\label{fig:tower10}
\end{figure}

\begin{figure}[!htbp]
	\includegraphics[width=320px]{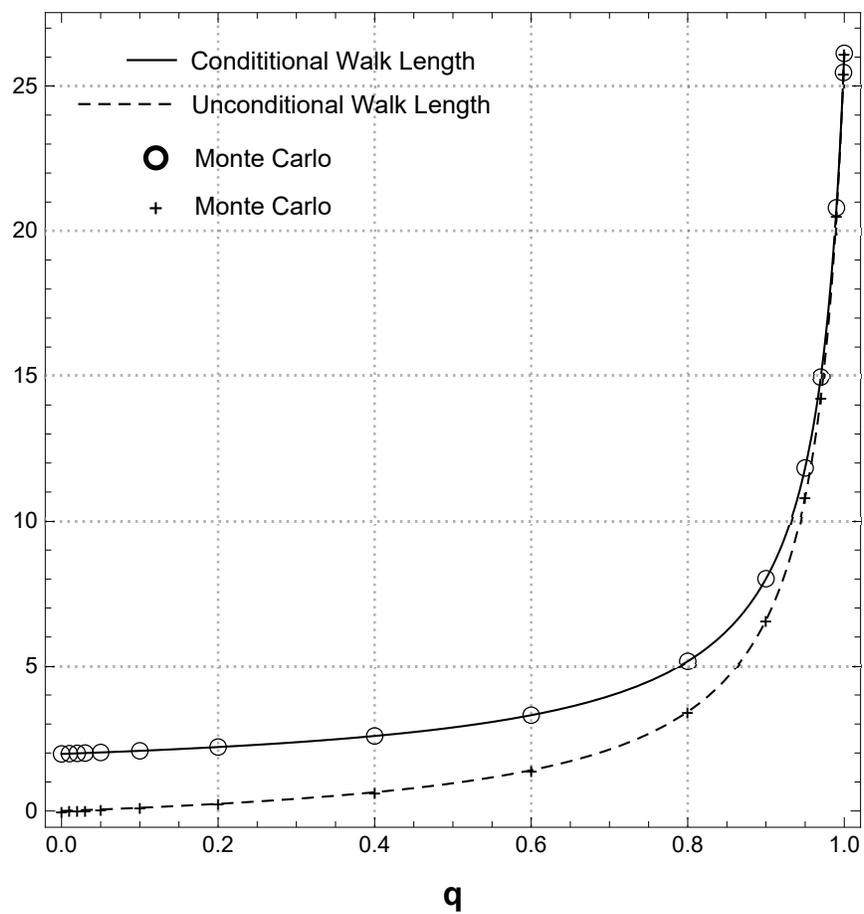}
	\caption{Mean walk length versus survival probability  $q$  for the $N=34$ Sierpinski tower (traps at
		three corner sites)}
	\label{fig:tower34}
\end{figure}

Displayed in figure \ref{fig:sierp15} are analytical and MC results for the overall conditional and unconditional walk length as a function of $q$ for the $N=15$ SG. Similar plots are presented in Fig. \ref{fig:sierp42} for the $N=42$ SG, in Fig. \ref{fig:tower10} for the $N=10$ Sierpinski tower, and in Fig. \ref{fig:tower34} for the $N=34$ Sierpinski tower. For the SG, traps are placed at two corner sites, whereas for the Sierpinski tower, traps are placed at three corner sites.

\FloatBarrier
Several conclusions can be drawn from these data. First, note the difference in the $q \to 0$ behavior between the unconditional walk length and the unconditional walk length (the latter quantity goes to an $N$-dependent, non-zero value in this limit). This point will be discussed in some detail in Section VI. Second, convergence to the classical random walk limit, $q \to 1$, is gradual for the conditional walk length, but less so for the unconditional walk length. Qualitatively, this means that, except for a narrow range of $q$ values, the lifetime of the diffusing particle becomes \emph{significantly} larger if one only counts absorption events at the deep trap than it is the case when all trajectories count, i.e., not only those terminated by absorption at the deep trap, but also by spontaneous death. Specifically, for $q \le 0.97$ there is an appreciable difference between conditional and unconditional walk lengths in all cases considered. Third, for sufficient large $q$, a walker persists much longer on larger gaskets than on smaller ones. The percentage of deep traps on the Sierpinski gasket is $33.3\%$ on the $N=6$ gasket, $13.3\%$ on the $N=15$ SG, $0.58\%$ on the $N=42$ SG, and $0.2\%$ on the $N=123$ SG. On the Sierpinski tower, the percentage of traps is $30.0\%$ on the $N=10$ tower, and $8.8\%$ on the $N=34$ tower. In contemporary language, the diffusing particle survives longer in "mesosystems" than in "nanosystems", with an appreciable difference between conditional and unconditional walk length in favor of the former as soon as $q$ falls below a relatively large (yet $N$-dependent) threshold value.

\section{Discussion}

As is evident from the results presented in the previous section, the difference between conditional and unconditional walk lengths is quite striking.  In order to decide which of these two scenarios is relevant in a given experimental problem, it is crucial to consider the $q \downarrow 0$  limit, where the difference is most pronounced.
 When $0 < q \ll 1$, or when $0 < q < 1$ but the generation
 number $n \gg 1$,  the random walk with the asymptotically leading contribution to $P_j(q)$ (the probability of a walker
 starting from $j$ reaching a trap) dominates. This  walk is not really random, but follows a directed shortest path from $j$ to the nearest trap. Since the number of steps in such a walk is equal to the number of time steps, we have $d_w = 1$ trivially, in this restricted case.

Occasionally, the starting site may be such that one or more traps are equidistant from that site via two or more paths.
For example,  on the Sierpinski gasket, the starting point (the top vertex, A) is equidistant from the traps at the two corner (bottom) vertices, L and R. In this case there are two paths, each of equal length, to the traps:
(i) 1 $\to$ 2 $\to$ straight down the side of the outer triangle, to L; and
(ii) 1 $\to$ 3 $\to$  straight down the other side of the  outer triangle, to R.  For a given generation number $n$,  the probabilities for each of these paths is $(q/2) (q/4)^{2^{n} - 1}$,
since the number of nearest neighbor (NN)  sites of  site A is just 2, while it is 4 for all the other sites on the walk till one reaches L or R.
But there are two  such walks. Hence,
\begin{equation}
P_1(q) \rightarrow q \left(\frac{q}{4}\right)^{2^{n}-1} = 4 \left(\frac{q}{4}\right)^{2^{n}},
\end{equation}
as stated earlier.

For every site, the leading contribution to $P_j(q)$
as $q \downarrow 0$
is easily found by identifying the shortest path to the nearest trap. The MFPT for that site is then just the number of bonds (or steps) on that path.
More formally, since the leading small-$q$ behavior of
$P_{j}(q)$  is a  monomial like $cq^{r}$ where $c$ is a constant,
 we have
 \begin{equation}
 T_{j} \rightarrow  q \left(\frac{d}{dq}\right) \ln \,(cq^{r})= r.
\end{equation}
Thus, while  both the probability distribution
$P_{j}$ and its first moment vanish as  $q \to 0$,
the ratio of the two quantities tends to a finite non-zero
value. Experimentally, if one measures the mean time to trapping for a walker with a very small survival probability $q$ at each step,
the particle flux impinging on the detector will be
much smaller than it would be in the case of diffusing
immortal particles,  because of the exponentially decreasing particle population. In applications where the instantaneous (or cumulative) flux is measured, the efficiency will be significantly lowered.
In the limit $q \to 0$ (large mortality) if what matters is the transit time of each arriving particle, then pseudo-ballistic transport will be observed as a result of an effective reduction of dimensionality. In other words, the mortality constraint selects the shortest (and therefore fastest) trajectories by penalizing long trajectories with a very low survival probability. This is a particular illustration of a more general feature reported previously, namely,  that
the statistical properties of the fraction of surviving particles in a collection of mortal walkers may be very different from those of an ensemble of standard random walkers \cite{AYL13, YAL14, M15}.

The  foregoing finding may be relevant for a number of experimental systems where diffusion-decay models inspired by Eq.
\eqref{diffeqnfundsoln}  fit experimental data very well. This is the case in a recent study \cite{CJ18},  where such a model was successfully used to reproduce the behavior of the detector signal in circular-dichroism experiments monitoring the valley diffusion current in TMDC quantum hetero-structures. Here, pairs of spin- and valley-polarized holes were first generated with a pulsed pump beam, and the density of valley-polarized holes was then measured at a different location by triggering a pulsed probe beam after a given time delay. The theoretical pump-probe signal was calculated by convoluting  the free solution of the diffusion-decay model with the spatial profile of the probe beam intensity. The resulting expression was used to fit experimental data and thereby obtain the values of the diffusion constant $D$ and
the valley lifetime. The value of the associated diffusion length
was found to be surprisingly large ($\sim 20\, \mu$s). In this case, holes that had already crossed the probe beam area before the pulse was triggered could still contribute to the circular dichroic detection signal if they happened to be revisiting that area at the time where the pulse had been triggered. Therefore, the computation of the detection signal is not a first-passage problem in this case, as opposed to the scenario considered above. However, to confirm the unusually large value of the diffusion length, one could envisage an alternative scenario in which the time needed by a hole to cover a certain distance is measured, whence the diffusion length can be inferred via Eq. \eqref{continuummfpt}.

In another recent work concerning anomalously large diffusion lengths, long-range exciton transport has been reported
\cite{XH18} in conjugated polymer nanofibers prepared by seeded growth. These nanostructures are assembled using a seeded-growth method for producing one- and two-dimensional templates of
controlled sizes \cite{XW07, HQ16}. In Ref. \cite{XH18}, Jin \emph{et al.} study exciton migration in crystalline fibers of poly(di-n-hexylfluorene) using photoluminescence quenching. They report exciton diffusion lengths greater than 200 nm, a significant increase (order of magnitude higher) than is realized in organic solar cells, and hence the significance of their work. Holmes \cite{RJH18}, in reviewing this work, notes that the results reported by Jin \emph{et al.} reinforce the idea that crystalline order (more precisely, a minimum degree of disorder) plays an important role in facilitating exciton diffusion. This work also shows that strong $p$-orbital overlap can as well enable more efficient exciton transport. Importantly, Jin \emph{et al.} conclude that these factors alone cannot fully explain the reported long diffusion lengths. Although Jin \emph{et al.} find their results compatible with a diffusion-decay model, Holmes concludes that to account for these new data, it is essential to recognize the limitations of diffusive or sub-diffusive transport regimes and to recognize the importance of ballistic transport.

We have explored in this paper the dramatic difference between the mean detection time of immortal particles and that of particles subjected to an exponential decay law diffusing on two fractal lattices, the Sierpinski gasket (fractal dimension $\approx 1.584$)
and the Sierpinski tower (fractal dimension $2$) embedded in Euclidean dimensions  $2$ and  $3$, respectively. We have illustrated that for a  $N=42$ Sierpinski gasket with traps at the lower two vertices, the exact value of the unconditional walk length is $428/5 = 85.6$, whereas for the conditional walk length it is $31/8 = 3.875$, a difference of more than an order of magnitude. From the point of view of first-passage properties, diffusion-decay models and ballistic behavior are not mutually exclusive, at least in some limit. For a given system, this means that it may be difficult to discern whether the behavior of certain quantities arises from pure ballistic behavior or from apparent ballistic behavior induced by a mortality constraint.  For the particular system considered in \cite{XH18}, this observation might help reconcile, to some extent,  the authors' conclusions with those in Holmes' commentary on their work. We suggest that this distinction can provide insight into studies of exciton transport in crystallization-driven  self-assembly of nanofibers templates,  and into the arrival properties and the statistics of detection of short-lived diffusing excitations present in other systems.

We close by emphasizing that the mechanism of reduction of dimensionality on which we have been focusing here relies on a mortality constraint; it is therefore fundamentally different from other mechanisms discussed in the literature, such as the one originally hypothesized by Adam and Delbruck \cite{AD68} and used as a source of inspiration in subsequent works (see, e.g., Refs. \cite{FGK76,LK84,MK92,KN11}). In the situation considered by Adam and Delbruck, the unconditional reaction rate of a diffusion-limited target search process is greatly enhanced, whereas in our case an increase in the conditional reaction rate is observed. That this is the case is remarkable given the decrease of the unconditional reaction efficiency due to the mortality of the diffusing species.

\section{Acknowledgements}

We dedicate this work to the memory of Gr\'egoire Nicolis.

E.A. thanks S.B. Yuste and A. Santos for stimulating discussions. The research of E.A. has been supported by the Spanish Agencia Estatal de Investigaci\'on through Grant No. FIS2016-76359-P and the Junta de Extremadura (Spain) through Grant No. GR18079, both partially financed by Fondo Europeo de Desarrollo Regional funds.

\end{document}